\documentclass[10pt,journal,compsoc]{IEEEtran}

\ifCLASSOPTIONcompsoc
  \usepackage[nocompress]{cite}
\else
  \usepackage{cite}
\fi

\ifCLASSINFOpdf
   \usepackage[pdftex]{graphicx}
   \graphicspath{{../}{../}}
   \DeclareGraphicsExtensions{.pdf,.jpeg,.png}
\else
   \usepackage[dvips]{graphicx}
   \graphicspath{{../eps/}}
   \DeclareGraphicsExtensions{.eps}
\fi

\usepackage{latexsym}
\usepackage{multirow}
\usepackage{isabelle,isabellesym}
\usepackage{amsmath}
\usepackage{amssymb}
\usepackage{color}
\usepackage{subfig}
\usepackage{float}
\usepackage{wasysym} 

\usepackage[colorlinks=true,citecolor=blue,linkcolor=blue,urlcolor=blue]{hyperref}
\usepackage{enumitem}

\newcommand{\skname}{ARINC SK}

\newtheorem{theorem}{Theorem}
\newtheorem{lemma}{Lemma}
\newtheorem{corollary}{Corollary}
\newtheorem{definition}{Definition}

\newcommand{\sectprefix}{{Section}}
\newcommand{\subsectprefix}{{Subsection}}
\newcommand{\figprefix}{{Fig.}}

\newcommand{\tableprefix}{{Table}}

\newcommand{\defprefix}{Definition}
\newcommand{\lemmaprefix}{Lemma}
\newcommand{\theoremprefix}{Theorem}

\newcommand{\tabincell}[2]{\begin{tabular}{@{}#1@{}}#2\end{tabular}}

\isabellestyle{it}

\let\endisabellecode=\endisabellecode

\newcommand{\isacodeftsz}{\footnotesize}  

\renewcommand\_{\textunderscore\allowbreak}

\newcommand{\interf}{\leadsto}

\newcommand{\sysconf}{conf}
\newcommand{\sched}{\mathbb{S}}
\newcommand{\transmitter}{\mathbb{T}}
\newcommand{\reachable}{\mathcal{R}}

\newcommand{\execution}[2]{{#1} \looparrowright {#2}}
\newcommand{\equidom}[3]{{#1}\stackrel{#2}{\sim}{#3}}
\newcommand{\equidomR}[3]{{#1}\stackrel{#2}{.\sim.}{#3}}
\newcommand{\equidomsub}[4]{{#1}\stackrel{#2}{\sim_{#3}}{#4}}
\newcommand{\equidomRsub}[4]{{#1}\stackrel{#2}{.\sim_{#3}.}{#4}}
\newcommand{\equidoms}[3]{{#1}\stackrel{#2}{\approx}{#3}}

\newcommand{\ssequidom}[3]{{#1}\stackrel{#2}{\asymp}{#3}}
\newcommand{\equivexec}[5]{\ssequidom{(\execution{#1}{#2})}{#3}{(\execution{#4}{#5})}}

\newcommand{\stateR}{\Psi}
\newcommand{\evtR}{\Theta}
\newcommand{\refine}{\sqsubseteq_{\stateR,\evtR}}

\newcommand{\zipbeforecode}{\vspace{6pt}}
\newcommand{\zipaftercode}{\vspace{6pt}}

\makeatletter
\newcommand{\superimpose}[2]{%
  {\ooalign{$#1\@firstoftwo#2$\cr\hfil$#1\@secondoftwo#2$\hfil\cr}}}
\makeatother
\newcommand{\ninterf}{\mathrel{\mathpalette\superimpose{{\slash}{\leadsto}}}}

\newcommand{\exist}{\checkmark}%
\newcommand{\halfexist}{\ast}

\begin{document}
%

\title{Refinement-based Specification and Security Analysis of Separation Kernels}


\author{Yongwang Zhao,
        David San\'an,
        Fuyuan Zhang,
        Yang Liu
\IEEEcompsocitemizethanks{\IEEEcompsocthanksitem This research is supported in part by the National Research Foundation, Prime Minister's Office, Singapore under its National Cybersecurity R\&D Program (Award No. NRF2014NCR-NCR001-30) and administered by the National  Cybersecurity R\&D Directorate. 
\IEEEcompsocthanksitem Y. Zhao is with the School of Computer Science and Engineering, Beihang Univerisity, Beijing, China (Email: zhaoyw@buaa.edu.cn) and the School of Computer Science and Engineering, Nanyang Technological University, Singapore (Email: ywzhao@ntu.edu.sg).
\IEEEcompsocthanksitem D. San\'an, F. Zhang and Y. Liu are with the School of Computer Science and Engineering, Nanyang Technological University, Singapore (Email: \{sanan, fuzh, yangliu\}@ntu.edu.sg).} 
}

\markboth{IEEE Transactions on Dependable and Secure Computing}%
{Shell \MakeLowercase{\textit{et al.}}: Bare Demo of IEEEtran.cls for Computer Society Journals}

\IEEEtitleabstractindextext{%
\begin{abstract}
Assurance of information-flow security by formal methods is mandated in security certification of separation kernels. 
As an industrial standard for improving safety, ARINC 653 has been complied with by mainstream separation kernels. Due to the new trend of integrating safe and secure functionalities into one separation kernel, security analysis of ARINC 653 as well as a formal specification with security proofs are thus significant for the development and certification of \underline{ARINC} 653 compliant \underline{S}eparation \underline{K}ernels ({\skname}s). 
This paper presents a specification development and security analysis method for {\skname}s based on refinement. We propose a generic security model and a stepwise refinement framework. Two levels of functional specification are developed by the refinement. A major part of separation kernel requirements in ARINC 653 are modeled, such as kernel initialization, two-level scheduling, partition and process management, and inter-partition communication. 
The formal specification and its security proofs are carried out in the Isabelle/HOL theorem prover. 
We have reviewed the source code of one industrial and two open-source {\skname} implementations, i.e. VxWorks 653, XtratuM, and POK, in accordance with the formal specification. During the verification and code review, six security flaws, which can cause information leakage, are found in the ARINC 653 standard and the implementations. 
\end{abstract}

}

\maketitle

\IEEEdisplaynontitleabstractindextext

\IEEEpeerreviewmaketitle

\IEEEraisesectionheading{\section{Introduction}\label{sect:intro}}
\IEEEPARstart{I}{n} recent years, a trend in embedded systems is to enable multiple applications from different vendors and with different criticality levels to share a common set of physical resources, such as IMA \cite{Parr99} and AUTOSAR \cite{AUTOSAR42}. To facilitate such a model, resources of each application must be protected from other applications in the system. The separation kernel \cite{Rushby81} and its variants (e.g. partitioning kernels and partitioning operating systems \cite{Rushby00,ARINC653p1}) establish such an execution environment by providing to their hosted applications spatial/temporal separation and controlled information flow. 
Separation kernels, such as PikeOS, VxWorks 653, INTEGRITY-178B, LynxSecure, and LynxOS-178, have been widely applied in domains from aerospace and automotive to medical and consumer electronics. 

Traditionally, safety and security of critical systems are assured by using two kinds of separation kernels respectively, such as VxWorks 653 for safety-critical systems and VxWorks MILS for security-critical systems. 
In order to improve the safety of separation kernels, the ARINC 653 standard \cite{ARINC653p1} has been developed to standardize the system functionality as well as the interface between the kernel and applications. ARINC 653 is the premier safety standard and has been complied with by mainstream separation kernels. On the other hand, the security of separation kernels is usually achieved by the Common Criteria (CC) \cite{CC} and Separation Kernel Protection Profile (SKPP) \cite{SKPP07} evaluation, in which formal verification of information-flow security is mandated for high assurance levels. 

A trend in this field is to integrate the safe and secure functionalities into one separation kernel. For instance, PikeOS, LynxSecure, and XtratuM are designed to support both safety-critical and security-critical solutions. Therefore, it is necessary to assure the security of the functionalities defined in ARINC 653 when developing \underline{ARINC} 653 compliant \underline{S}eparation \underline{K}ernels ({\skname}s). Moreover, a security verified specification compliant with ARINC 653 and its mechanically checked proofs are highly desirable for the development and certification of {\skname}s. 
The highest assurance level of CC certification (EAL 7) requires comprehensive security analysis using formal representations of the security model and functional specification of {\skname}s as well as formal proofs of correspondence between them.
Although formal specification \cite{craig07,velykis10,Verb14,Klaus15} and verification \cite{Heitmeyer08,Richards10,Wilding10,freit11,Murray13,Dam13,sanan14} of information-flow security on separation kernels have been widely studied in academia and industry, information-flow security of {\skname}s has not been studied to date. To the best of our knowledge, our work is the first effort on this topic in the literature. 

There exist three major challenges to be addressed in this work. 
First, the ARINC 653 standard is highly complicated. It specifies the system functionality and 57 standard services of separation kernels using more than 100 pages of informal descriptions. 
Second, as a sort of hyperproperties \cite{Clarkson10}, it is difficult to automatically verify information-flow security on separation kernels so far. Formal analysis of information-flow security of separation kernels is difficult and needs exhausting manual efforts. 
Moreover, there exist different definitions of information-flow security (e.g. in \cite{rushby92,sabelfeld03,von04,Murray12}) and the relationship of them on {\skname}s has to be clarified. 
Third, reusability of formal specification and proofs is important. They should be extensible and easy to be reused for subsequent development and certification. 

This paper presents a specification development and security analysis method based on stepwise refinement \cite{Wirth71,Back89} for {\skname}s with regard to the EAL 7 of CC certification. 
A separation kernel is an event-driven system reacting to hypercalls or exceptions, which is alternatively called a reactive system. Therefore, we borrow design elements from existing formalisms such as superposition refinement \cite{Back96} of reactive systems and Event-B \cite{Abrial07}. 
We start from a generic security model of {\skname}s that could be instantiated as functional specifications at different abstract levels. The refinement in this paper concretizes an abstract functional specification by incorporating additional design elements and by transforming existing ones. The stepwise refinement addresses the first challenge by modeling complicated requirements of ARINC 653 in a modular and hierarchical manner. 
The refinement provides formal proofs of the correspondence between functional specifications for CC certification. 
Information-flow security is proven on an abstract specification and it is preserved in a concrete specification by means of refinement. 
Moreover, system functionalities and services of ARINC 653 are modeled as the event specification in the functional specification. 
In such a design, information-flow security is proven in a compositional way, i.e. the security of {\skname}s is implied by the satisfaction of local properties (\emph{unwinding conditions}) on events. Thus, the second challenge is resolved. 
Using the correctness-by-construction method based on the refinement, well-structured specification of {\skname}s is developed together with their correctness and security proofs, which helps to resolve the third challenge. 

The challenges mentioned above are not well addressed in the literature. 
In formal verification of the seL4 microkernel \cite{Klein14} and its separation extension \cite{Murray13}, and the ED separation kernel \cite{Heitmeyer08}, refinement methods have been applied. First, due to the post-hoc verification objective of these projects, refinement is not a technique to develop the specification in a stepwise manner, but to prove the conformance between formalizations at different levels. Therefore, they have few levels of specification and the refinement is coarse-grained. Second, ARINC 653 is not the emphasis of these works and the formal specification are not compliant with ARINC 653. 
Information-flow security verification has been enforced on PikeOS \cite{Verb14,Klaus15}, INTEGRITY-178B \cite{Richards10}, and an ARM-based separation kernel \cite{Dam13}. However, refinement is not considered in these works. 
Correctness-by-construction methods have been used to create formal specification of separation kernels \cite{craig07,velykis10,freit11,zhao15}. However, information-flow security is not the emphasis of them. 
Formalization and verification of ARINC 653 have been studied in recent years, such as the formal specification of ARINC 653 architecture \cite{Oliveira12}, modeling ARINC 653 for model-driven development of IMA applications \cite{Delan10}, and formal verification of application software on top of ARINC 653 \cite{de11}. In \cite{zhao15}, the system functionalities and all service in ARINC 653 have been formalized in Event-B, and some inconsistencies have been found in the standard. 
These works aim at the safety of separation kernels and applications. Our work is the first to conduct a formal security analysis of ARINC 653.

We have used Isabelle/HOL \cite{Nipkow02} to formalize the security model, the refinement method, and functional specifications as well as to prove information-flow security. All specifications and security proofs are available at ``\url{http://securify.scse.ntu.edu.sg/skspecv3/}''. 
In detail, the technical contributions of this work are as follows.

\begin{enumerate}
\item \label{contr:1} We define a security model, which is a parameterized abstraction for the execution and security configuration of {\skname}s. 
A set of information-flow security properties are defined in the model. An inference framework is proposed to sketch out the implications of the properties and an unwinding theorem for compositional reasoning. 
\item \label{contr:2} We propose a refinement framework for stepwise development of {\skname}s. Functional specifications of {\skname}s at different levels are instantiations of the security model, and thus all properties of the security model are satisfied on the specifications. We define a security extended superposition refinement for {\skname}s, which supports introducing additional design elements (e.g. new state variables and new events). We also show the security proofs of the refinement, i.e. the preservation of security properties during the refinement. 
\item \label{contr:3} We develop a top-level specification for {\skname}s which covers kernel initialization, partition scheduling, partition management, and inter-partition communication (IPC) according to ARINC 653. 
A second-level specification is developed by refining the top-level one. We add processes, process scheduling, and process management according to ARINC 653. Security is proven by refinement. 
\item \label{contr:4} We conduct a code-to-spec review required by CC certification on one industrial and two open-source {\skname} implementations, i.e. VxWorks 653, XtratuM \cite{xtratum}, and POK \cite{pok}, in accordance with the formal specification. 
During the verification and code review, six covert channels to leak information \cite{millen99} have been found in the ARINC 653 standard and the implementations. The approaches to fixing them are also provided. 
\end{enumerate}

In this paper, we have extended our previous work \cite{Zhao16} by introducing the security model, the refinement framework, the second-level specification, the code review of VxWorks 653, and four new covert channels. 
The rest of this paper is organized as follows. {\sectprefix} \ref{sect:prelim} introduces preliminaries of this paper. {\sectprefix} \ref{sect:approach} presents the overview of our method. The next four sections present the security model, refinement framework, top-level and second-level specifications, respectively. Then, {\sectprefix} \ref{sect:reslt_disc} discusses the security flaws found in ARINC 653 and the implementations. Finally, {\sectprefix} \ref{sect:concl} gives the conclusion and future work.

\section{Preliminaries}
\label{sect:prelim}

\subsection{Information-flow Security}
The notion \emph{noninterference} is introduced in \cite{Goguen82} in order to provide a formal foundation for the specification and analysis of information-flow security policies. 
The idea is that a security domain $u$ is noninterfering with a domain $v$ if no action performed by $u$ can influence the subsequent outputs seen by $v$. 
Language-based information-flow security \cite{sabelfeld03} assigns either \emph{High} or \emph{Low} labels to program variables and ensures the data confidentiality by preventing information leakage from \emph{High}-level data to \emph{Low}-level data. 
Transitive noninterference is too strong and thus is declassified as intransitive one \cite{Mantel04,Krohn09}. That is, the system should allow certain flows of information from \emph{High} domains to \emph{Low} domains, if that flow traverses the appropriate declassifier. In \cite{rushby92}, intransitive noninterference is defined in a state-event manner and concerns the visibility of \emph{events}, i.e. the secrets that events introduce in the system state. 
Language-based information-flow security is generalized to arbitrary multi-domain policies in \cite{von04} as a new state-event based notion \emph{nonleakage}. In \cite{von04}, nonleakage and intransitive noninterference are combined as a new notion \emph{noninfluence}, which considers both the data confidentiality and the secrecy of events. 
Intransitive noninterference \cite{rushby92,von04,Ramirez14} in state-event manner has been applied to verify separation kernels, such as seL4 \cite{Murray13}, PikeOS \cite{Klaus15}, INTEGRITY-178B \cite{Richards10}, and AAMP7G \cite{Wilding10}.

\subsection{ARINC 653}
The ARINC 653 standard is organized in six parts. Part 1 \cite{ARINC653p1} in Version 3 specifies the baseline operating environment for application software used within IMA. 
It defines the \emph{system functionality} and requirements of 57 \emph{services}. 
{\skname}s are mandated to comply with this part. The six major functionalities specified in ARINC 653 are partition management, process management, time management, inter- and intra-partition communication, and health monitoring. 

ARINC 653 establishes and separates multiple partitions in time and space except the controlled information flows along communication channels among partitions. 
The security policy used by {\skname}s is the \emph{Inter-Partition Flow Policy} (IPFP) \cite{Levin07}, which is intransitive. It is expressed abstractly in a partition flow matrix $\mathbf{partition\_flow}: partition \times partition \rightarrow mode$, whose entries indicate the mode of the flow. For instance, $\mathbf{partition\_flow}(P_1,P_2) = QUEUING$ means that partition $P_1$ is allowed to send information to partition $P_2$ via a channel with a message queue. 
Another feature of ARINC 653 is its two-level scheduling, i.e. partition scheduling and process scheduling. For the purpose of temporal separation, the partition scheduling in ARINC 653 is a fixed, cycle based scheduling and is strictly deterministic over time. Process scheduling in a partition is priority preemptive.

\subsection{Isabelle/HOL}
Isabelle is a generic and tactic-based theorem prover. We use Isabelle/HOL \cite{Nipkow02}, an implementation of high-order logic in Isabelle, for our development. The keyword \textbf{datatype} is used to define an inductive data type. 
The list in Isabelle is one of the essential data type in computing and is defined as $\textbf{datatype} \ 'a \ list = Nil \ (``[]") \ | \ Cons \ 'a \ ``\ 'a \ list \ " (infixr \ ``\#")$, where $[\ ]$ is the empty list and $\#$ is the concatenation. 
The polymorphic option type is defined as $\textbf{datatype} \ 'a \ option = None \ | \ Some \ (the: \ 'a)$. A function of type $'b \rightharpoonup \ 'a$ models a partial function, which is equal to $'b \Rightarrow ('a \ option)$.  Record types may be defined, for example $\textbf{record} \ point = xcoord::int \ \ ycoord::int$ with the elements like $p = {\isasymlparr} xcoord = 10, ycoord = 10 {\isasymrparr}$ and projections $xcoord \ p$ and $ycoord \ p$. Records may be updated, such as $p{\isasymlparr} xcoord := 20, ycoord := 20 {\isasymrparr}$, and extended, such as $\textbf{record} \ cpoint = point + col :: colour$. 
Nonrecursive definitions can be made with the \textbf{definition} command and the \textbf{primrec} function definition is used for primitive recursions. 
\emph{Locales} are Isabelle's approach for dealing with parametric theories. Locales may be instantiated by assigning concrete data to parameters, and the resulting instantiated declarations are added to the current context. This is called \emph{locale interpretation}. In its simplest form, a \textbf{locale} declaration consists of a sequence of context elements declaring parameters (the keyword \textbf{fixes}) and assumptions (the keyword \textbf{assumes}).

To enhance readability, we will use standard mathematical notation where possible. Examples of formal specification will be given in the Isabelle/HOL syntax. 

\section{Method Overview}
\label{sect:approach}

\begin{figure}[t]
\centerline{\includegraphics[width=3.4in]{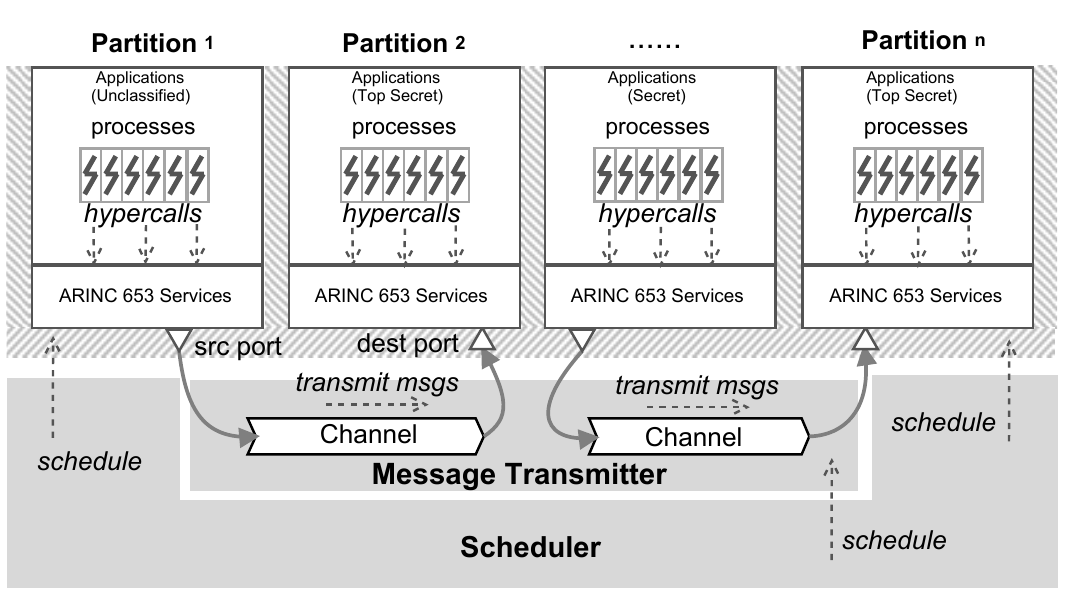}}
\caption{Architecture of ARINC 653 Separation Kernels}
\label{fig:arch}
\end{figure}

The architecture of {\skname}s we consider in this paper is shown in {\figprefix} \ref{fig:arch}. Since ARINC 653 Part 1 in Version 3 \cite{ARINC653p1} is targeted at single-core processing environments, we consider single-core separation kernels. For simplicity, separation kernels usually disable interrupts in kernel mode \cite{Murray13} and thus there is no in-kernel concurrency, which means that the hypercalls and system events (e.g. scheduling) are executed in an atomic manner. 

We adopt intransitive noninterference in state-event manner due to its practicality in industry and intransitive channel-control policies used in ARINC 653. 
We formalize various definitions of information-flow security, e.g. noninterference, nonleakage, and noninfluence. 
An inference framework of these properties is provided, in which the implication relationship among the properties is proven. 

Since automatic analysis of information-flow security is still restricted by the size of state space (\cite{Eggert11,Clarkson14}), it is difficult to deal with operating system kernels so far. 
Automatic analysis techniques (e.g. model checking) usually verify one configuration of the system at a time \cite{Ha04}. However, the deployment of partitions on separation kernels is unknown in advance. 
Therefore, it is well suited to use theorem proving based approaches. Interactive theorem proving requires human intervention and creativity, but produces machine-checkable proofs required by CC certification. Moreover, it is not constrained to specific properties or finite state spaces. This approach is mostly adopted to formal verification of operating systems \cite{Klein09b}. 
We choose Isabelle/HOL in this work because (1) it can deal with large-scale formal verification of real-world systems \cite{Andron12} due to the expressiveness of HOL, a high degree of proof automation, and powerful libraries, etc.; (2) most of the OS verification are using Isabelle/HOL, e.g. seL4 \cite{Klein14,Murray13} and PikeOS \cite{Verb14,Klaus15}; and (3) Isabelle/HOL is recommended as a formal methods tools as required by CC certification on high assurance levels \cite{Blas15}. Although other approaches, e.g. Event-B, provide the refinement framework, properties of information-flow security could not be formulated straightforwardly by their inherent specification languages. We could see some efforts on this direction \cite{Hoang13} but no supported tools.

\begin{figure}[t]
\centerline{\includegraphics[width=3.2in]{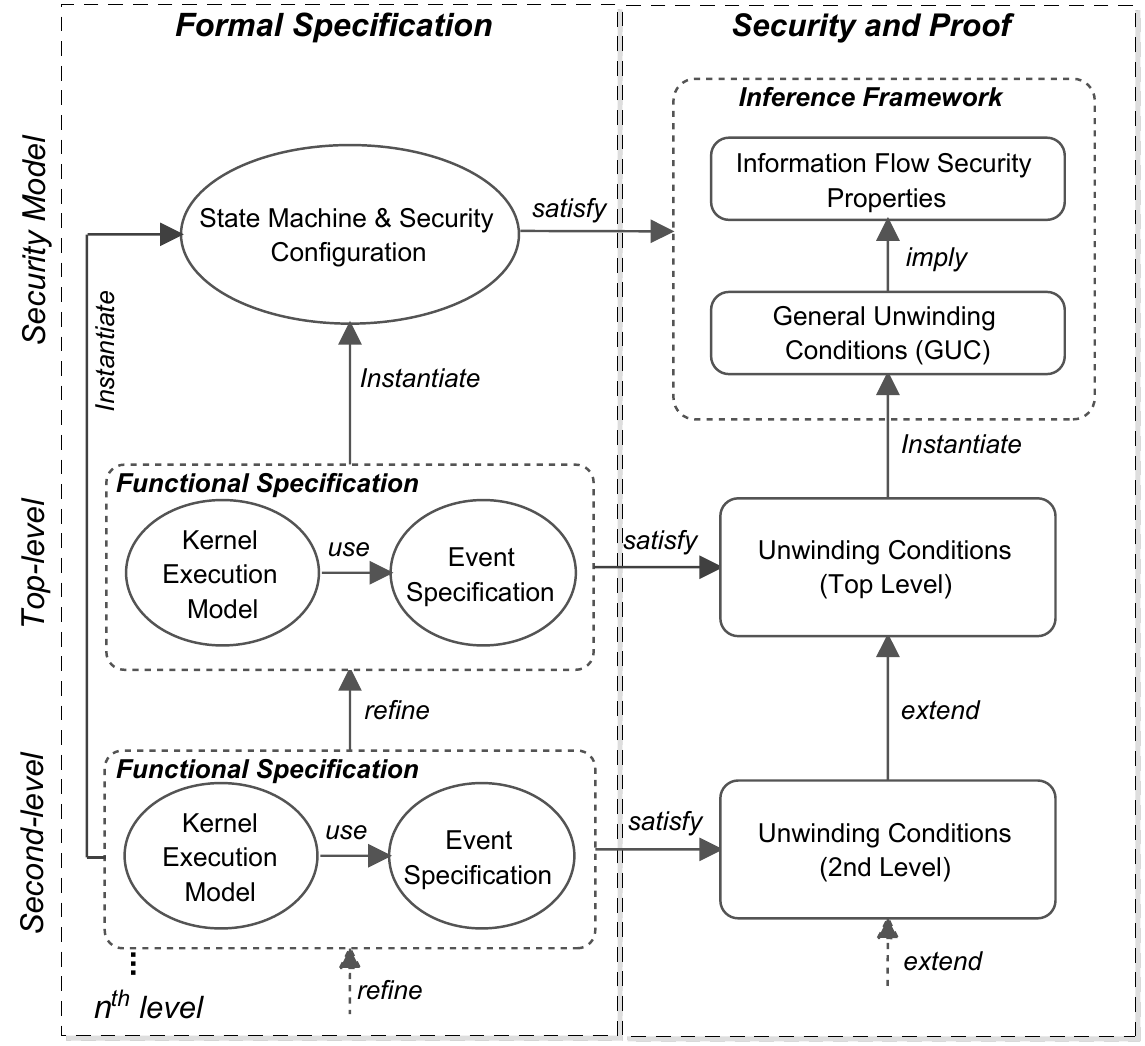}}
\caption{Method Overview}
\label{fig:method}
\end{figure}

The method overview is shown in {\figprefix} \ref{fig:method}. 
The security model is a generic model of {\skname}s for information-flow security, which includes a state machine based execution model and the security configuration. The inference framework presents the implication among these properties and an unwinding theorem. It follows the classical unwinding relation based proof \cite{rushby92}, i.e. satisfaction of the general unwinding conditions (GUCs) implies the security. 
The security model is a parameterized abstraction, which means that the kernel components are defined as abstract types. It could be instantiated to a concrete model, i.e. a functional specification of {\skname}s. By this instantiation, all elements (e.g. properties, implications, and proofs) of the security model are reused and preserved in the specification. 
In order to improve the reusablity of the specification and proofs, a functional specification is decomposed into two parts: a kernel execution model and an event specification for ARINC 653. The execution model is an instance of the security model, while the event specification defines Isabelle/HOL functions to implement the state changes when an event occurs. The concrete functions in the event specification are invoked by the execution model. By the instantiation, the properties of the inference framework are preserved on the specification. The rest of security proofs of the specification is to show satisfaction of unwinding conditions (UCs) which are instances of GUCs. 

In order to support additional design elements in the refinement, we use superposition refinement \cite{Back96} in this work. Since formal specification of {\skname}s combines the system dynamics and security components, we extend the superposition refinement with security constraints for {\skname}s. The state represented by a set of state variables at the abstract level is transformed into concrete state variables and extended by new state variables. An event is refined to a set of concrete events and new events may be introduced in the refinement. Therefore, in addition to a state simulation relation, we use an event relation to connect the abstract and concrete events. 
To show information-flow security at the concrete level, UCs of the abstract level are extended due to the new state variables and new events. The proofs of UCs at the concrete level only consider the new state variables and security proofs at the abstract level are reused. 
In the case of data refinement, i.e. without introducing new events and new state variables, security is automatically preserved in the refinement. 

A major part of requirements in ARINC 653 are modeled in this work. 
We develop two levels of specification by refinement. {\skname}s use IPC to implement controlled information flows among partitions. Moreover, in the IPFP policy of ARINC 653, communication ports and channels are associated with partitions, and all processes in a partition can access the ports configured for this partition. Therefore, we model functionalities related to information flows, i.e. partition management and IPC services, in the top-level functional specification.
For functional completeness, kernel initialization and partition scheduling are also modeled. At the second level, we use the refinement to add process scheduling and process management services, which are functionalities in partitions. Other functionalities and services of ARINC 653 are possible to be added by refinement in future.

\section{Security Model of Separation Kernels}
\label{sect:gen_model}
We design a security model for {\skname}s in this section. The model consists of a nondeterministic state machine, the security configuration of {\skname}s, information-flow security properties, and an inference framework of these properties. 

\subsection{State Machine and Security Configuration}
The state-event based information-flow security uses a state machine to represent the system model. For universality, we adopt a nondeterministic model. The state machine of {\skname}s is defined as follows.

\begin{definition}[State Machine of {\skname}s]
$\mathcal{M} = \langle \mathcal{S}, \mathcal{E}, \varphi, s_0 \rangle$ is a tuple, where $\mathcal{S}$ is the state space, $\mathcal{E}$ is the set of event labels, $s_0 \in \mathcal{S}$ is the initial state, and $\varphi: \mathcal{E} \rightarrow \mathbb{P}(\mathcal{S} \times \mathcal{S})$ is the state-transition function. 
\end{definition}

The $\varphi$ function characterises the single-step behaviour of separation kernels by executing an event, such as a hypercall or scheduling. The auxiliary functions used in the security model are defined in detail in {\figprefix} \ref{fig:aux_funs}. Based on the $\varphi$ function, we define the $run$ function to represent the execution of a sequence of events. 
The $execution(s,es)$ function (denoted as $\execution{s}{es}$) returns the set of final states by executing a sequence of events $es$ from a state $s$, where $\lhd$ is the domain restriction of a relation. By the $execution$ function, the reachability of a state $s$ is defined as $reachable(s)$ (denoted as $\mathcal{R}(s)$). 

\begin{figure}[t]
\hspace*{-0.5cm}   
{
\footnotesize
\begin{tabular}{ll} 
$ 
\left\{
\begin{aligned}
& run([\ ]) = Id \\
& run(e\#es) = \varphi(e) \circ run(es)
\end{aligned}
\right.
$ 
&
$ 
\begin{aligned}
& \execution{s}{es} \triangleq \{s\} \lhd run(es) \\
& \mathcal{R}(s) \triangleq  \exists es. \ (s_0,s)\in run(es)
\end{aligned}
$ 
\vspace{5pt}
\\
\multicolumn{2}{l}{
$\equidoms{s}{D}{t} \triangleq \forall d \in D.\ \equidom{s}{d}{t}$
\ \ \ \ \ 
$\ssequidom{ss}{d}{ts} \triangleq \forall s \ t. \ s \in ss \wedge t \in ts \longrightarrow \equidom{s}{d}{t}$
}
\vspace{5pt}
\\ 
\multicolumn{2}{l}{
$ 
\left\{
\begin{aligned}
& sources ([ \ ], s, d) = \{d\} \\
& sources (e \# es, s, d) = (\bigcup \{sources (es, s', d) \mid (s,s') \in \varphi(e)\}) \ \cup \\
& \quad \quad \quad \quad \quad \quad \quad \quad \quad \quad \quad \{w \mid w = kdom(s,e) \wedge (\exists v \ s'. \ (w \interf v) \\
& \quad \quad \quad \quad \quad \quad \quad \quad \quad \quad \quad \wedge (s,s')\in \varphi(e) \wedge v \in sources(es,s',d))\}

\end{aligned}
\right.
$ 
}
\vspace{5pt}
\\
\multicolumn{2}{l}{
$ 
\left\{
\begin{aligned}
& ipurge ([ \ ], d, ss) = [ \ ] \\
& ipurge (e \# es, d, ss) = \mathbf{if} (\exists s \in ss. \ kdom(s,e) \in sources(e \# es, s, d)) \\
& \quad \quad \quad \quad \quad \quad \quad \quad \quad \quad \mathbf{then} \\
& \quad \quad \quad \quad \quad \quad \quad \quad \quad \quad \quad e \# ipurge (es, d, (\bigcup_{s \in ss}\{s' \mid (s,s') \in \varphi(e)\})) \\
& \quad \quad \quad \quad \quad \quad \quad \quad \quad \quad \mathbf{else} \ \ ipurge (es, d,ss) 
\end{aligned}
\right.
$ 
}
\end{tabular}
}
\caption{Auxiliary Functions in Security Model}
\label{fig:aux_funs}
\end{figure}

The security configuration of separation kernels is usually comprised of security domains, security policies, domain of events, and state equivalence as follows. 

\textbf{Security domains}: Applications in partitions have various security levels. We consider each partition as a security domain. Since partition scheduling is strictly deterministic over time, we define a security domain \emph{scheduler} for partition scheduling, which cannot be interfered by any other domains to ensure that the \emph{scheduler} does not leak information via its scheduling decisions. Note that process scheduling is conducted in partitions. Since ARINC 653 defines the channel-based communication services using ports and leaves the implementation of message transmission over channels to underlying separation kernels, a security domain \emph{message transmitter} is defined for message transmission. The transmitter also decouples message transmission from the scheduler to ensure that the scheduler is not interfered by partitions. 
Therefore, the security domains ($\mathcal{D}$) of {\skname}s are the scheduler ($\sched$), the transmitter ($\transmitter$), and the configured partitions ($\mathcal{P}$), i.e. $\mathcal{D}=\mathcal{P}\cup \{\sched, \transmitter\}$.

\textbf{Security policies}: In order to discuss information-flow security policies, we assume an \emph{interference relation} $\interf \subseteq \mathcal{D} \times \mathcal{D}$ according to the $\mathbf{partition\_flow}$ matrix and $\ninterf$ is the complement relation of $\interf$. If there is a channel from a partition $p_1$ to a partition $p_2$, then $p_1 \interf \transmitter$ and $\transmitter \interf p_2$ since the transmitter is the message intermediator. Since the scheduler can possibly schedule other domains, it can interfere with them. In order to prevent the scheduler from leaking information via its scheduling decisions, we require that no other domains can interfere with the scheduler.

\textbf{Domain of events}: Traditional formulations in the state-event based approach assume a static mapping from events to domains, such that the domain of an event can be determined solely from the event itself \cite{rushby92}. However, in separation kernels that mapping is dynamic \cite{Murray12}. When a \emph{hypercall} occurs, the kernel must consult the partition scheduler to determine which partition is currently executing, and the current executing partition is the domain of the hypercall. On the other hand, separation kernels can provide specific hypercalls which are only available to privileged partitions. Therefore, the domain of an event is represented as a partial function $kdom: \mathcal{S} \times \mathcal{E} \rightharpoonup \mathcal{D}$.

\textbf{State equivalence}: It means states are \emph{identical} for a security domain. We define an equivalence relation $\sim$ on states for each domain such that $\equidom{s}{d}{t}$ if and only if states $s$ and $t$ are identical for domain $d$, that is to say states $s$ and $t$ are indistinguishable for $d$. 
Two states are equivalent for the scheduler when domains of each event in the two states are the same. 
For a set of domains $D$, we define $\equidoms{s}{D}{t}$ as shown in {\figprefix} \ref{fig:aux_funs} to represent that states $s$ and $t$ are equivalent for all domains in $D$. For two sets of states $ss$ and $ts$, we define $\ssequidom{ss}{d}{ts}$ as shown in {\figprefix} \ref{fig:aux_funs} to represent that any two states in $ss$ and $ts$ are equivalent for domain $d$.

Based on the discussion above, we define the security model of {\skname}s as follows. We assume that each event is always enabled in a state. Whenever an event $e$ should not be executed in a state $s$, it does not change the state. 

\begin{definition}[Security Model of {\skname}s]
\label{def:sec_model}
\[\mathcal{SM} = \langle \mathcal{M}, \mathcal{D}, kdom, \interf, \sim \rangle\] with assumptions as follows.
\begin{enumerate}
\item $\forall p_1, p_2 \in \mathcal{P}. \ p_1 \interf p_2 \longrightarrow p_1 \interf \transmitter \wedge \transmitter \interf p_2$
\item $\forall d \in \mathcal{D}. \ \sched \interf d$
\item $\forall d \in \mathcal{D}. \ d \interf \sched \longrightarrow d = \sched$
\item $\sim$ is an equivalence relation.
\item $\forall s \ t \ e. \ \equidom{s}{\sched}{t} \longrightarrow kdom(s,e) = kdom(t,e)$
\item events are enabled in any state, i.e. $\forall s \ e. \ \mathcal{R}(s) \longrightarrow (\exists s'. \ (s,s') \in \varphi(e))$. 
\end{enumerate}
\end{definition}

The security model is represented in Isabelle/HOL as a locale $\mathcal{SM}$, which could be instantiated to create functional specification. 

\subsection{Information-flow Security Properties}
 
In order to express the allowed information flows for the intransitive policies, we use a function $sources(es,s,d)$ as shown in {\figprefix} \ref{fig:aux_funs}, which yields the set of domains that are allowed to pass information to domain $d$ when event sequence $es$ occurs from state $s$. It is inductively defined on event sequences. We include in $sources(e\#es, s, d)$ all domains that can pass information to $d$ when $es$ occurs from all successor states $s'$ of $s$, as well as the domain $kdom(s,e)$ performing the event $e$, whenever there exists some intermediate domain $v$ by which the domain $kdom(s,e)$ can pass information to $d$ indirectly.
In the intransitive purged sequence ($ipurge(es,d,ss)$ in {\figprefix} \ref{fig:aux_funs}), the events of partitions that are not allowed to pass information to $d$ directly or indirectly are removed. Given an event sequence $e\#es$ executing from a set of state $ss$, $ipurge$ keeps the first event $e$ if this event is allowed to affect the target domain $d$, i.e. $\exists s \in ss. \ kdom(s,e) \in sources(e \# es, s, d)$. It then continues on the remaining events $es$ from the successor states of $ss$ after executing $e$.  On the other hand, if $e$ is not allowed to affect the target domain $d$, then it is removed from the sequence and purging continues on the remaining events $es$ from the current set of states $ss$.
The observational equivalence of an execution is denoted as  $\equivexec{s}{es_1}{d}{t}{es_2}$, which means that a domain $d$ is identical to any two final states after executing $es_1$ from $s$ ($\execution{s}{es_1}$) and executing $es_2$ from $t$. 

\begin{table}[t]
\centering
\footnotesize 

\caption{Information-flow Security Properties} 
\begin{tabular} {|l|l|}
\hline
\textbf{No}. & \textbf{Properties}
\\ \hline 
\multirow{2}*{(1)} & \textbf{noninterference}:
\\
& $\forall d\ es.\ \equivexec{s_0}{es}{d}{s_0}{ipurge(es,d,\{s_0\})}$
\\ \hline
\multirow{2}*{(2)} & \textbf{weak\_noninterference}:
\\
& 
$\begin{aligned}
\forall d\ es_1\ es_2.\ & ipurge(es_1,d,\{s_0\}) = ipurge(es_2,d,\{s_0\}) \\
&\longrightarrow (\equivexec{s_0}{es_1}{d}{s_0}{es_2})
 \end{aligned}$
\\ \hline
\multirow{2}*{(3)} & \textbf{noninterference\_r}:
\\
& 
$\forall d\ es\ s.\ \reachable(s)
\longrightarrow (\equivexec{s}{es}{d}{s}{ipurge(es,d,\{s\}})$
\\ \hline
\multirow{2}*{(4)} & \textbf{weak\_noninterference\_r}:
\\
& 
$\begin{aligned}
\forall d\ es_1\ es_2\ & s.\ \reachable(s) \\
& \wedge ipurge(es_1,d,\{s\}) = ipurge(es_2,d,\{s\}) \\
& \longrightarrow (\equivexec{s}{es_1}{d}{s}{es_2})
 \end{aligned}$
\\ \hline
\multirow{2}*{(5)} & \textbf{nonleakage}:
\\
& 
$\begin{aligned}
\forall d\ es\ s\ t.& \ \reachable(s) \wedge \reachable(t) \wedge (\equidom{s}{\sched}{t}) 
\wedge (\equidoms{s}{sources(es,s,d)}{t}) \\
& \longrightarrow (\equivexec{s}{es}{d}{t}{es})
 \end{aligned}$
\\ \hline
\multirow{2}*{(6)} & \textbf{weak\_noninfluence}:
\\
& 
$\begin{aligned}
\forall & d\ es_1\ es_2\ s\ t. \ \reachable(s) \wedge \reachable(t) \wedge (\equidoms{s}{sources(es_1,s,d)}{t}) \\
& \wedge (\equidom{s}{\sched}{t}) \wedge ipurge(es_1, d, \{s\}) = ipurge(es_2, d, \{t\}) \\
& \longrightarrow (\equivexec{s}{es_1}{d}{t}{es_2})
 \end{aligned}$
\\ \hline
\multirow{2}*{(7)} & \textbf{noninfluence}:
\\
& 
$\begin{aligned}
\forall d\ es\ s\ t. &\ \reachable(s) \wedge \reachable(t) \wedge (\equidoms{s}{sources(es,s,d)}{t}) \wedge (\equidom{s}{\sched}{t}) \\
& \longrightarrow (\equivexec{s}{es}{d}{t}{ipurge(es, d, \{t\})})
 \end{aligned}$
\\ \hline
\end{tabular}
\label{tbl:props}
\end{table}

The intransitive noninterference \cite{rushby92} on the execution model is defined as \emph{noninterference} in {\tableprefix} \ref{tbl:props}. Its intuitive meaning is that events of domains that cannot interfere with a domain $d$ should not affect $d$. Formally, given an event sequence $es$ and a domain $d$, final states after executing $es$ from the initial state $s_0$ and after executing its purged sequence from $s_0$ are identical to $d$. 
In \emph{noninterference}, the $ipurge$ function only deletes all unsuitable events. Another version is introduced in \cite{von04} to handle arbitrary insertion and deletion of secret events, which is defined as \emph{weak\_noninterference} in {\tableprefix} \ref{tbl:props}. In \emph{weak\_noninterference}, we only ask for two event sequences which have the same effects on the target domain $d$, i.e. $ipurge(es_1,d,\{s_0\}) = ipurge(es_2,d,\{s_0\})$. 
The above definitions of noninterference are based on the initial state $s_0$, but separation kernels usually support \emph{warm} or \emph{cold start} and they may start to execute from a non-initial state. Therefore, we define a more general version \emph{noninterferece\_r} based on the reachable function $\mathcal{R}$. This general noninterference requires that systems starting from any reachable state are secure. It is obvious that this noninterference implies the classical noninterference due to $\reachable(s_0) = True$. \emph{Weak\_noninterference} is also generalized as a new property \emph{weak\_noninterference\_r}.

Nonleakage and noninflunce are defined for {\skname}s based on the scheduler as shown in {\tableprefix} \ref{tbl:props}. 
The intuitive meaning of nonleakage is that if data are not leaked initially, data should not be leaked during executing a sequence of events. 
Separation kernels are said to preserve nonleakage when for any pair of reachable states $s$ and $t$ and an observing domain $d$, if (1) $s$ and $t$ are equivalent for all domains that may (directly or indirectly) interfere with $d$ during the execution of $es$, i.e. $\equidoms{s}{sources(es, s, d)}{t}$, and (2) the same domain is currently executing in both states, i.e. $\equidom{s}{\sched}{t}$, then $s$ and $t$ are observationally equivalent for $d$ and $es$. Noninfluence is the combination of nonleakage and noninterference to ensure that there is no secrete data leakage and secrete events are not visible according to security policies. 
Murray et al. \cite{Murray12} have defined noninfluence for seL4, which is a weaker one and defined as \emph{weak\_noninfluence} in {\tableprefix} \ref{tbl:props}. Similar to \emph{weak\_noninterference}, it can handle arbitrary insertion and deletion of secret events. It is the combination of \emph{weak\_noninterference\_r} and \emph{nonleakage}. We propose a stronger one, i.e. \emph{noninfluence}, according to the original definition \cite{von04} by extending the scheduler and the state reachability. It is the combination of \emph{noninterference\_r} and \emph{nonleakage}. In the next subsection, we will show that noninfluence is actually the conjunction of noninterference and nonleakage.

\subsection{Inference Framework of Security Properties}
\label{subsec:inf_frm}

\begin{figure}[t]
\centerline{\includegraphics[width=3.0in]{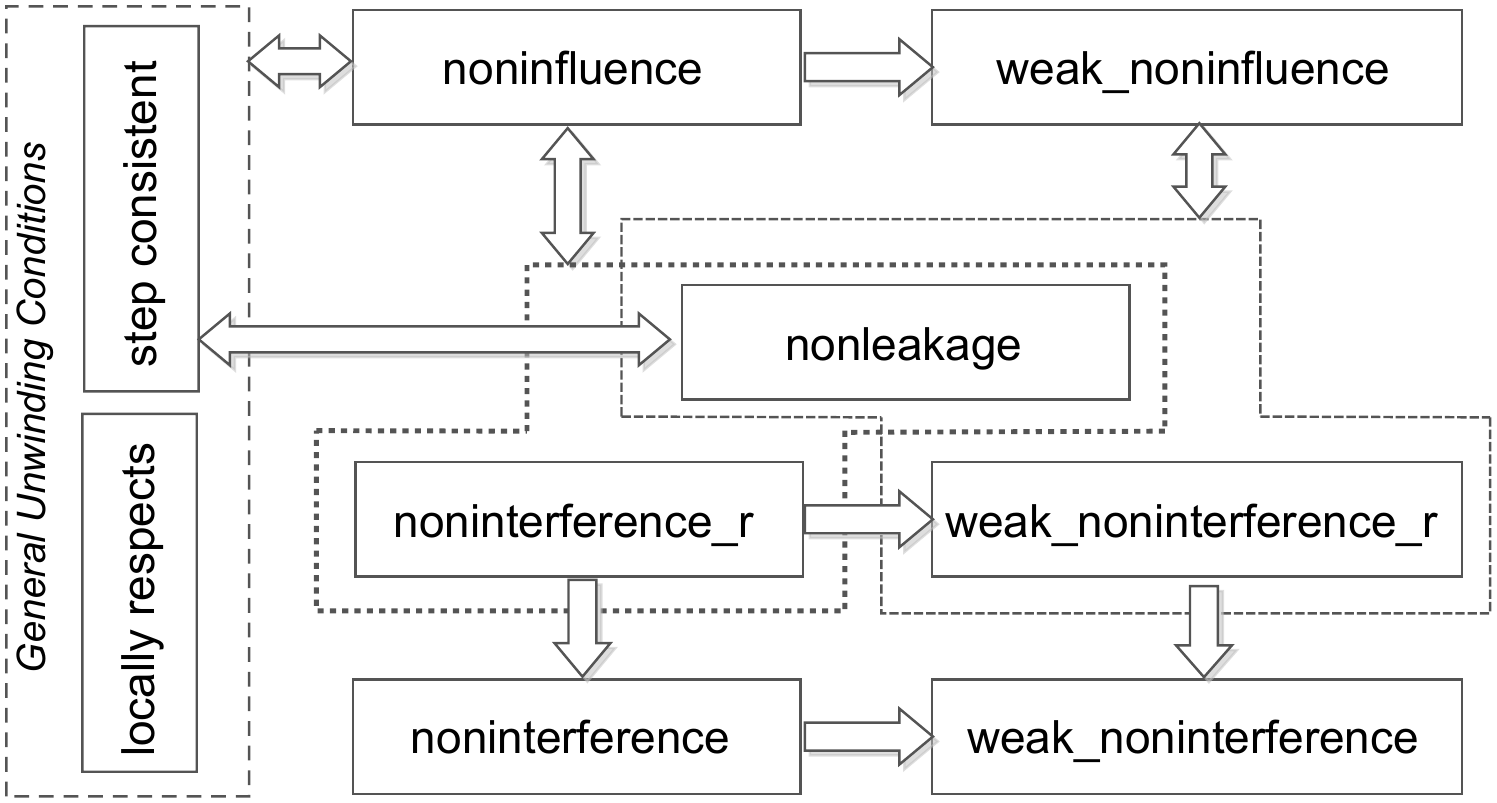}}
\caption{Inference Framework of Security Properties}
\label{fig:infer_frm}
\end{figure}

The inference framework clarifies the implication relations among the security properties on the security model. We have proven all implication relations as shown in {\figprefix} \ref{fig:infer_frm}, where an arrow means the implication between properties. 

The standard proof of information-flow security is discharged by proving a set of unwinding conditions \cite{rushby92} that examine individual execution steps of the system. This paper also follows this approach. 
We first define the unwinding conditions on event (UCE) as follows.

\begin{definition}[Step Consistent Condition of Event]
\label{def:sc_e}
\end{definition}
{
\small
\[
\begin{aligned}
SC(e) & \triangleq \forall d\ s\ t. \ \reachable(s) \wedge \reachable(t) \wedge (\equidom{s}{d}{t}) \wedge (\equidom{s}{\sched}{t})\\
& \wedge (kdom(s,e) \interf d) \wedge (\equidom{s}{kdom(s,e)}{t}) \\
& \longrightarrow (\forall s' \ t'. \ (s,s') \in \varphi(e) \wedge (t,t') \in \varphi(e) \longrightarrow \equidom{s'}{d}{t'})
\end{aligned}
\]
}

\begin{definition}[Locally Respects Condition of Event]
\label{def:lr_e}
\end{definition}
{
\small
\[
\begin{aligned}
LR(e) \triangleq \forall d\ s\ s'.\ & \reachable(s)\ \wedge (kdom(s,e) \ninterf d) \wedge (s,s')\in \varphi(e) \\
& \longrightarrow (\equidom{s}{d}{s'})
\end{aligned}
\]
}

The \emph{step consistent} condition requires that for any pair of reachable states $s$ and $t$, and any observing domain $d$, the next states after executing an event $e$ in $s$ and $t$ are indistinguishable for $d$, i.e. $\equidom{s'}{d}{t'}$, if (1) $s$ and $t$ are indistinguishable for $d$, (2) the current executing domains in $s$ and $t$ are the same, (3) the domain of $e$ in state $s$ can interference with $d$, and (4) $s$ and $t$ are indistinguishable for the domain of $e$.
The \emph{locally respects} condition means that an event $e$ that executes in a state $s$ can affect only those domains to which the domain executing $e$ is allowed to send information. 

Based on the UCEs, the general unwinding conditions of the security model can be defined as 
$SC \triangleq \forall e. \ SC(e)$ and $LR \triangleq \forall e. \ LR(e)$. 
The soundness and completeness of unwinding conditions in the security model are shown as follows, where the soundness is the unwinding theorem in \cite{rushby92,von04}.

\begin{theorem}[Soundness and Completeness of Unwinding Conditions]
\label{thm:uw_theorem}
The security model $\mathcal{SM}$ satisfies that
\end{theorem}
{
\small
\begin{equation*}
(SC \wedge LR) = noninfluence \ \textbf{and} \ SC = nonleakage
\end{equation*}
}
\vspace{-10pt}

The objective of noninfluence is to ensure data confidentiality and secrecy of events by combing noninterference and nonleakage. We show the soundness and completeness of noninfluence in the security model as follows.

\begin{theorem}[Soundness and Completeness of Noninfluence]
\label{thm:sound_cmplt_noninf}
The security model $\mathcal{SM}$ satisfies that
\end{theorem}
{
\small
\begin{equation*}
\begin{aligned}
& noninfluence = (noninterference\_r \wedge nonleakage) \ \textbf{and} \\ 
& 
\begin{aligned}
weak\_noninfluence = & \ (weak\_noninterference\_r \\
& \wedge nonleakage)
\end{aligned}
\end{aligned}
\end{equation*}
}
\vspace{-16pt}


\section{Specification and Refinement Framework}

After discussing the security model, we present the functional specification by instantiating the security model and a refinement framework for stepwise specification development in this section. 

\subsection{Specification by Security Model Instantiation}
A functional specification consists of a kernel execution model and an event specification. The kernel execution model is an instance of the security model.
The instantiation is an interpretation of the $\mathcal{SM}$ locale, i.e.
$
SM_I = \mathbf{interpretation} \ \mathcal{SM}\{\allowbreak s_0 / s_{0I}, \allowbreak \varphi / \varphi_I, \allowbreak  kdom / kdom_I, \allowbreak \sched/Sched, \allowbreak \transmitter/Trans, \allowbreak \sim/\sim_I, \allowbreak \interf/\interf_I \}
$, where the parameters of $\mathcal{SM}$ are substituted by concrete ones. In order to assure the correctness of the instantiation, we provide the \emph{instantiation proof} to show that the instance parameters preserve the assumptions of $\mathcal{SM}$. 

The event specification is a mapping from events to a set of functions to model the functionalities and services in ARINC 653. The $\varphi_I$ function in the execution model executes events by invoking these functions. 
Thus, a functional specification of {\skname}s is defined as follows. 

\begin{definition} [Functional Specification of {\skname}s]
\label{def:spec}
A functional specification is a tuple $SK = \langle SM_I, F\rangle$, where 
\begin{enumerate}
\item $SM_I = \langle \mathcal{M}_I, \mathcal{D}_I, kdom_I, \interf_I, \sim_I \rangle$ is the execution model, where $\mathcal{M}_I = \langle \mathcal{S}_I, \mathcal{E}_I, \varphi_I, s_{0I} \rangle$. 
\item $F: \mathcal{E}_I \rightarrow \Gamma$ is the event specification, where $\Gamma$ is the set of functions and each $f \in \Gamma$ has the type $\mathcal{S}_I \rightarrow \mathbb{P}(\mathcal{S}_I)$.
\item $\varphi_I$ in $\mathcal{M}_I$ defines the execution of each event $e$, where $\varphi_I(e) = \{(s,t) \mid t \in F(e)(s)\}$.
\end{enumerate}
\end{definition}

\subsection{Specification Refinement}

In the superposition refinement, existing state variables and events can be refined. In addition, new state variables and events can be added to an abstract specification. Here, we use subscripts $A$ and $C$ to represent the abstract and concrete specifications, respectively. During the refinement the abstract state $\mathcal{S}_A$ can be transformed into $\mathcal{S}_T$ and new state variables denoted as $\mathcal{S}_{\Delta}$ can be incorporated. Thus, the concrete state $\mathcal{S}_C = \mathcal{S}_T + \mathcal{S}_{\Delta}$, where the symble ``+'' is the extension of data type (e.g. \textbf{record} extension in Isabelle/HOL). The security extended superposition refinement for the functional specification is defined as follows.

\begin{definition}[Refinement]
\label{def:refine}
Given two functional specifications $SK_A$ and $SK_C$, $SK_C$ refines $SK_A$ using a state simulation relation $\stateR: \mathcal{S}_C \rightarrow \mathcal{S}_A$ and an event relation $\evtR: \mathcal{E}_C \rightarrow (\mathcal{E}_A \cup \{\tau\})$, denoted as $SK_A \refine SK_C$, if
\begin{enumerate}
\item the initial state in $SK_C$ establishes $R$, i.e. $\stateR(s_{0C}) = s_{0A}$.
\item each event in $\mathcal{E}_A$ is refined by a set of events in $\mathcal{E}_C$, i.e. $\evtR$ is a surjection and $\forall e\ s\ t. \ \evtR(e) \neq \tau \wedge (s,t) \in \varphi_C(e) \longrightarrow (\stateR(s),\stateR(t)) \in \varphi_A(\evtR(e))$. 
\item new events only change the new state variables introduced in the refinement and does not affect the variables in $SK_A$, i.e. $\forall e\ s\ t. \ \evtR(e) = \tau \wedge (s,t) \in \varphi_C(e) \longrightarrow \stateR(t) = \stateR(s)$.
\item security domains are preserved and refined events has the same execution domain as that at the abstract level, i.e. $\mathcal{D}_C = \mathcal{D}_A$ and $\forall e\ s. \ \evtR(e) \neq \tau \longrightarrow kdom_C(s,e) = kdom_A(\stateR(s),\evtR(e))$. 
\item the refinement does not change the interference relation, i.e. $\interf_C \ = \ \interf_A$. 
\item $\equidomsub{s}{d}{C}{t} = \equidomsub{\stateR(s)}{d}{A}{\stateR(t)} \wedge \equidomsub{s}{d}{\Delta}{t}$, where $\sim_{\Delta}$ only considers the new state variables $\mathcal{S}_{\Delta}$ introduced in the refinement. 
\end{enumerate}
\end{definition}

The relation $\stateR$ maps concrete states to abstract ones. It is actually the transformation of $S_A$ to $S_T$. Abstract states of next states by executing an event $e$ in a state $t$ at the concrete level is a subset of next states by executing the refined event $\evtR(e)$ in a state $s$ ($s = \stateR(t)$) at the abstract level. If the event is a new one at the concrete level, its execution should not affect the abstract state. We could consider that a new event at the concrete level refines a $\tau$ action, which does not change the abstract states. The security configuration at the abstract level is preserved in the refinement, i.e. (1) $\sched$, $\transmitter$, partitions $\mathcal{D}$, and the relation $\interf$ are the same, (2) the domain of events at the abstract level is preserved, and (3) the state equivalence relation at the abstract level is preserved, while the relation at the concrete level also requires that the two states are equivalent for the new state variables (i.e. $\sim_{\Delta}$). 

The refinement in this paper is reflexive and transitive. 
The correctness is ensured by {\theoremprefix} \ref{thm:refine_sound}. Since noninterference considers the state and event together, our refinement implies that the state-event trace set of a concrete specification is a subset of the abstract one. Thus, the refinement preserves noninterference as we present in the next subsection. 

\begin{theorem}[Soundness of Refinement]
\label{thm:refine_sound}
If $SK_A$ and $SK_C$ are instances of $\mathcal{SM}$ and $SK_A \refine SK_C$, then
{
\begin{equation*}
\forall es. \ \stateR^*(\execution{s_{0C}}{es}) \subseteq \execution{s_{0A}}{\evtR^*(es)}
\end{equation*}
}
\end{theorem}

In the theorem, $\stateR^*$ maps a set of concrete states to a set of abstract states by the relation $\stateR$, and $\evtR^*$ maps a sequence of concrete events to a sequence of abstract events by the relation $\evtR$. The $\tau$ event is ignored during executing a sequence of events at the abstract level. 

From the definition of the refinement, we could see that for any reachable state $s$ at the concrete level, its abstract state on the $\stateR$ relation is also reachable at the abstract level.

\begin{lemma}[State Reachability in Refinement]
\label{lemma:reach_refine}
If $SK_A$ and $SK_C$ are instances of $\mathcal{SM}$ and $SK_A \refine SK_C$, then
{
\begin{equation*}
\forall s . \ \mathcal{R}_C(s) \longrightarrow \mathcal{R}_A(\stateR(s))
\end{equation*}
}
\end{lemma}

Superposition refinement \cite{Back96} assumes the termination of new events introduced in the refinement. Since hypercalls and system events of {\skname}s terminate, we assume that events terminate in our refinement framework. In the formal specification of {\skname}s, we use the \textbf{primrec} and \textbf{definition} in Isabelle/HOL to define the event specification. Thus, the termination of events is automatically ensured in Isabelle/HOL.

\subsection{Security Proofs of Refinement}
The security proofs of a concrete specification also follow the unwinding theorem introduced in {\subsectprefix} \ref{subsec:inf_frm}. Since a concrete specification is an instance of the security model, we have following theorem according to {\theoremprefix} \ref{thm:uw_theorem}.

\begin{theorem}[Unwinding Theorem for Concrete Specification]
\label{thm:uw_thm_concrete}
If $SK_C$ is an instances of $\mathcal{SM}$ and $SK_C$ satisfies $SC_C \wedge LR_C$, then $SK_C$ satisfies $noninfluence$.
\end{theorem}

We define the step consistent condition of events on the new state variables as follows. 
\begin{definition}[Step Consistent Condition of Event on New State Variables]
\end{definition}
{\small
\[
\begin{aligned}
S&C_{C\Delta} (e) \triangleq \forall \ d\ s\ t. \ \reachable_C(s) \wedge \reachable_C(t) \wedge (\equidomsub{s}{d}{C}{t}) \wedge (\equidomsub{s}{\sched}{C}{t})\\
& \wedge (kdom_C(s,e) \interf_C d) \wedge (\equidomsub{s}{kdom_C(s,e)}{C}{t}) \\
& \longrightarrow (\forall s' \ t'. \ (s,s') \in \varphi_C(e) \wedge (t,t') \in \varphi_C(e) \longrightarrow \equidomsub{s'}{d}{\Delta}{t'})
\end{aligned}
\]
}

The locally respects condition of events on the new state variables is defined in an analogous manner. Based on definitions of the refinement and the unwinding conditions as well as {\lemmaprefix} \ref{lemma:reach_refine}, we have four lemmas shown as follows.

\begin{lemma}[Step Consistent of Event Refinement]
\label{lemma:sc_e_ref}
If $SK_A$ and $SK_C$ are instances of $\mathcal{SM}$ and $SK_A \refine SK_C$, then
\end{lemma}
{\small
\begin{equation*}
\forall \ e . \ \evtR(e) \neq \tau \wedge SC_A(\evtR(e)) \wedge SC_{C\Delta}(e) \longrightarrow SC_C(e)
\end{equation*}
\vspace{-16pt}
}

\begin{lemma}[Locally Respects of Event Refinement]
\label{lemma:lr_e_ref}
If $SK_A$ and $SK_C$ are instances of $\mathcal{SM}$ and $SK_A \refine SK_C$, then
\end{lemma}
{\small
\begin{equation*}
\forall \ e . \ \evtR(e) \neq \tau \wedge LR_A(\evtR(e)) \wedge LR_{C\Delta}(e) \longrightarrow LR_C(e)
\end{equation*}
\vspace{-16pt}
}

\begin{lemma}[Step Consistent of New Event in Refinement]
\label{lemma:sc_ne_ref}
If $SK_A$ and $SK_C$ are instances of $\mathcal{SM}$ and $SK_A \refine SK_C$, then
\end{lemma}
{\small
\begin{equation*}
\forall \ e . \ \evtR(e) = \tau \wedge SC_{\Delta}(e) \longrightarrow SC_C(e)
\end{equation*}
\vspace{-16pt}
}

\begin{lemma}[Locally Respects of New Event in Refinement]
\label{lemma:lr_ne_ref}
If $SK_A$ and $SK_C$ are instances of $\mathcal{SM}$ and $SK_A \refine SK_C$, then
\end{lemma}
{\small
\begin{equation*}
\forall \ e . \ \evtR(e) = \tau \wedge LR_{\Delta}(e) \longrightarrow LR_C(e)
\end{equation*}
\vspace{-16pt}
}

Finally, based on these lemmas and {\theoremprefix} \ref{thm:uw_thm_concrete}, we have {\theoremprefix} \ref{thm:sec_conc_spec}. The theorem means that if $SK_C$ is a refinement of $SK_A$ and $SK_A$ satisfies the unwinding conditions, we only need to prove the unwinding conditions on new state variables to show information-flow security of $SK_C$. 

\begin{theorem}[Security of Refinement]
\label{thm:sec_conc_spec}
If $SK_A$ and $SK_C$ are instances of $\mathcal{SM}$, $SK_A$ satisfies $SC_A \wedge LR_A$, $SK_C$ satisfies $SC_{\Delta} \wedge LR_{\Delta}$, and $SK_A \refine SK_C$, then $SK_C$ satisfies $noninfluence$.
\end{theorem}

In the case of data refinement, i.e. without new events and new state variables introduced in the refinement, information-flow security is automatically preserved on the concrete specification. 

\begin{corollary}
\label{cor:sec_conc_spec_ne}
If $SK_A$ and $SK_C$ are instances of $\mathcal{SM}$, $SK_A$ satisfies $SC_A \wedge LR_A$, $SK_A \refine SK_C$, $\nexists e. \ \evtR(e) = \tau$, and $\forall s \ t \ d. \ \equidomsub{s}{d}{\Delta}{t}$, then $SK_C$ satisfies $noninfluence$. 
\end{corollary}

\section{Top-level Specification and Security Proofs}
In the top-level functional specification, we model kernel initialization, partition scheduling, partition management, and inter-partition communication defined in ARINC 653. We first instantiate the security model as the kernel execution model. Then, we present the event specification. Finally, security proofs are discussed.

\subsection{Kernel Execution Model}
Basic components at the top level are partitions and communication objects. 
A \emph{partition} is basically the same as a program in a single application environment \cite{ARINC653p1}. IPC is conducted via messages on channels configured among partitions. Partitions have access to channels via \emph{ports} which are the endpoints of channels. The modes of transferring messages over channels are \emph{queuing} and \emph{sampling}. 
A significant characteristic of {\skname}s is that the basic components are statically configured at built-time. 
Partitions, communication objects, and the system configuration are specified in Isabelle/HOL as ``\textbf{record} \textit{Sys\_Config}'' and defined as a constant ``\emph{conf::Sys\_Config}'' in the top-level specification. 
When creating ports by invoking IPC services in a partition, only configured ports in the partition are created. 
A set of configuration constraints are defined to ensure the correctness of the system configuration. 

We first instantiate the security model by a set of concrete parameters as follows. 

\textbf{Events}:
We consider two types of events in the specification: \emph{hypercalls} and \emph{system events}. Hypercalls cover all partition management and IPC services in ARINC 653. System events are the actions of the kernel itself and include kernel initialization, scheduling, and transmitting messages over channels. Other types of events could be introduced during subsequent refinements. Events are illustrated in {\figprefix} \ref{fig:arch} as dotted line arrows and italics. 

\textbf{Kernel State and Transition}:
In the top-level specification, the kernel state concerns states of partitions, the scheduler, and ports. The state of a partition consists of its operating mode ($partitions$) and the created ports ($part\_ports$). The state of the scheduler shows which is the current executing partition ($cur$). The state of a port is mainly about messages in its buffer ($comm$). The $\textbf{datatype} \ Port\_Type$ models sampling and queuing mode ports as well as their message buffers. 
We define the type of $port\_id$ as natural number and use the partial functions to store the created ports and their states. We prove a set of invariants about types in the specification, such as the domain of $part\_ports$ is the same as that of $comm$ in all reachable states.

\begin{isabellec}
\isacodeftsz
\zipbeforecode
\ \ \ \ \isacommand{record}\isamarkupfalse%
\ State\ {\isacharequal}\ partitions\ {\isacharcolon}{\isacharcolon}\ part\_id\ {\isasymrightharpoonup}\ part\_mode\_type\isanewline
\ \ \ \ \ \ \ \ \ \ \ \ \ \ \ \ \ \ \ \ \ \ \ \ \ \ \
part\_ports\ {\isacharcolon}{\isacharcolon}\ port\_id\ {\isasymrightharpoonup}\ part\_id\isanewline
\ \ \ \ \ \ \ \ \ \ \ \ \ \ \ \ \ \ \ \ \ \ \ \ \ \ \
cur\ {\isacharcolon}{\isacharcolon}\ domain\_id\ \ \ \ \ \ 
comm\ {\isacharcolon}{\isacharcolon}\ port\_id\ {\isasymrightharpoonup}\ Port\_Type
\zipaftercode
\end{isabellec}

The state transition function $\varphi$ is instantiated as the $exec\_event$ function in the top-level specification, in which the mapping function $F$ in {\defprefix} \ref{def:spec} is also modeled. 

\textbf{Event Domain}:
The domain of the system events is static: the domain of the event \emph{scheduling} is $\sched$ and the domain of \emph{message transmission} is $\transmitter$. The domain of hypercalls is dynamic and depends on the current state of the kernel. It is defined as $kdom\ s\ (\mathbf{hyperc}\ h) = cur\ s$, where $cur\ s$ returns the current executing partition in the state $s$. 

\textbf{Interference Relation}:
The interference relation in the security model is instantiated as the function \emph{interference1} as follows in the top-level specification. 

\begin{isabellec}
\isacodeftsz
\zipbeforecode
\isacommand{definition}\isamarkupfalse%
\ interference1\ {\isacharcolon}{\isacharcolon}\ {\isachardoublequoteopen}domain{\isacharunderscore}id\ {\isasymRightarrow}\ domain{\isacharunderscore}id\ {\isasymRightarrow}\ bool{\isachardoublequoteclose}\ {\isacharparenleft}{\isachardoublequoteopen}{\isacharparenleft}{\isacharunderscore}\ {\isasymleadsto}\ {\isacharunderscore}{\isacharparenright}{\isachardoublequoteclose}{\isacharparenright}\isanewline
\ \ \ \isakeyword{where}\ {\isachardoublequoteopen}interference{\isadigit{1}}\ d{\isadigit{1}}\ d{\isadigit{2}}\ {\isasymequiv}\isanewline
\ \ \ \ \ \ \textbf{if}\ d{\isadigit{1}}\ {\isacharequal}\ d{\isadigit{2}}\ \textbf{then}\ True\isanewline
\ \ \ \ \ \ \textbf{else\ if}\ is{\isacharunderscore}sched\ conf\ d{\isadigit{1}}\ \textbf{then}\ True\ \isanewline
\ \ \ \ \ \ \textbf{else\ if}\ {\isasymnot}{\isacharparenleft}is{\isacharunderscore}sched\ conf\ d{\isadigit{1}}{\isacharparenright}\ {\isasymand}\ {\isacharparenleft}is{\isacharunderscore}sched\ conf\ d{\isadigit{2}}{\isacharparenright}\ \textbf{then}\ False\isanewline
\ \ \ \ \ \ \textbf{else\ if}\ is{\isacharunderscore}part\ conf\ d{\isadigit{1}}\ {\isasymand}\ is{\isacharunderscore}trm\ conf\ d{\isadigit{2}}\ \textbf{then}\ part{\isacharunderscore}intf{\isacharunderscore}trm\ conf\ d{\isadigit{1}}\isanewline
\ \ \ \ \ \ \textbf{else\ if}\ is{\isacharunderscore}trm\ conf\ d{\isadigit{1}}\ {\isasymand}\ is{\isacharunderscore}part\ conf\ d{\isadigit{2}}\ \textbf{then}\ trm{\isacharunderscore}intf{\isacharunderscore}part\ conf\ d{\isadigit{2}}\isanewline
\ \ \ \ \ \ \textbf{else}\ False{\isachardoublequoteclose}
\zipaftercode
\end{isabellec}

The $interference1$ is in conformance with the assumptions 1) - 3) in {\defprefix} \ref{def:sec_model}. 

\textbf{State Equivalence}:
For a partition $d$, $\equidom{s}{d}{t}$ if and only if $vpeq\_part \ s \ d \ t$, which is defined as follows.

\begin{isabellec}
\isacodeftsz
\zipbeforecode
\isacommand{definition}\isamarkupfalse%
\ vpeq\_part\ {\isacharcolon}{\isacharcolon}\ ``State\ {\isasymRightarrow}\ part\_id\ {\isasymRightarrow}\ (State\ {\isasymtimes}\ bool)''\ \isakeyword{where}\isanewline
\ \ vpeq\_part\ s\ d\ t\ {\isasymequiv}
 (partitions\ s)\ d\ =\ (partitions\ t)\ d {\isasymand} vpeq\_part\_comm\ s\ d\ t
\zipaftercode
\end{isabellec}

It means that states $s$ and $t$ are equivalent for partition $d$, when the partition state and the communication abilities of $d$ on these two states are the same. 
An example of the communication ability is that if a destination queuing port $p$ is not empty in two states $s$ and $t$, partition $d$ has the same ability on $p$ in $s$ as in $t$. This is because $d$ is able to receive a message from $p$ in the two states.
The equivalence of communication abilities defines that partition $d$ has the same set of ports, and that the number of messages is the same for all destination ports on states $s$ and $t$. 

Two states $s$ and $t$ are equivalent for the scheduler when the current executing partition in the two states are the same. The equivalence of states for the transmitter requires that all ports and states of the ports are the same. 

\subsection{Event Specification}
\label{subsect:evt_spc_top}
The event specification on the top level models kernel initialization, partition scheduling, all services of IPC and partition management defined in ARINC 653. 

\textbf{Kernel Initialization and Scheduling}:
The kernel initialization considers initialization of the kernel state, which is defined as $s_0 = system\_init \ \sysconf$. The $system\_init$ function assigns initial values to the kernel state according to the kernel configuration. 
Because the execution of message transmission is also under the control of scheduling, we define an abstract partition scheduling as follows that non-deterministically chooses one partition or the transmitter as the current executing domain, where \emph{partconf} is a field of \emph{Sys\_Config} with the type $partition\_id \rightharpoonup Partition\_Conf$ to store the configured partitions.

\begin{isabellec}\isacodeftsz
\zipbeforecode
\isacommand{definition}\isamarkupfalse%
\ schedule\ {\isacharcolon}{\isacharcolon}\ ``Sys\_Config\ {\isasymRightarrow}\ State\ {\isasymRightarrow}\ State''\ \isakeyword{where} \isanewline 
\ \ \ \ \ schedule\ sc\ s\ {\isasymequiv}\ s{\isasymlparr}cur \ :=\ SOME\ p{\isachardot} p{\isasymin}  {\isacharbraceleft}x{\isachardot}\ {\isacharparenleft}partconf\ sc{\isacharparenright}\ x\ {\isasymnoteq}\ None\isanewline
\; \; \; \; \; \; \; \; \; \; \; \; \; \; \; {\isasymor}\ x\ {\isacharequal}\ $\transmitter$ sc{\isacharbraceright}{\isasymrparr}
\zipaftercode
\end{isabellec}

\textbf{Partition Management}:
Partition management services in ARINC 653 are available to the application software for setting a partition's operating mode and to obtain a partition's status. 
The \emph{Set\_Partition\_Mode} service request is used to set the operating mode of the current partition to \emph{NORMAL} after the initialization of the partition is complete. The service is also used for setting the partition back to \emph{IDLE} (partition shutdown), and to \emph{COLD\_START} or \emph{WARM\_START} (partition restart), when a fault is detected and processed. 
The \emph{Set\_Partition\_Mode} service is specified as follows. The partition mode transition in ARINC 653 does not allow transitting from \emph{COLD\_START} to \emph{WARM\_START}. The function updates the operating mode of the current executing partition. 

\begin{isabellec}
\isacodeftsz
\zipbeforecode
\isacommand{definition}\isamarkupfalse%
\ set\_part\_mode\ {\isacharcolon}{\isacharcolon}\ {\isachardoublequoteopen}Sys\_Config\ {\isasymRightarrow}\ State\ {\isasymRightarrow}\ part\_mode\_type\ {\isasymRightarrow}\ State{\isachardoublequoteclose}\ \isakeyword{where}\isanewline
\ \ {\isachardoublequoteopen}set\_part\_mode\ sc\ s\ m\ {\isasymequiv}\ \isanewline
\ \ \ \ \ \ {\isacharparenleft}\textbf{if}\ {\isacharparenleft}partitions\ s{\isacharparenright}\ {\isacharparenleft}cur\ s{\isacharparenright}\ {\isasymnoteq}\ None\ {\isasymand}\isanewline
\ \ \ \ \ \ \ \ \ \ {\isasymnot}\ {\isacharparenleft}the\ {\isacharparenleft}{\isacharparenleft}partitions\ s{\isacharparenright}\ {\isacharparenleft}cur\ s{\isacharparenright}{\isacharparenright}\ {\isacharequal}\ COLD\_START \isanewline
\ \ \ \ \ \ \ \ \ \ \ \ \ \ \ {\isasymand}\ m\ {\isacharequal}\ WARM\_START{\isacharparenright}\ \textbf{then}\isanewline
\ \ \ \ \ \ \ \ \textbf{let}\ pts\ {\isacharequal}\ partitions\ s \textbf{in}
\ s{\isasymlparr}partitions\ {\isacharcolon}{\isacharequal}\ pts{\isacharparenleft}cur\ s\ {\isacharcolon}{\isacharequal}\ Some\ m {\isacharparenright}{\isasymrparr}\isanewline
\ \ \ \ \ \ \textbf{else} \ s{\isacharparenright}\isanewline
\ \ {\isachardoublequoteclose}
\zipaftercode
\vspace{-6pt}
\end{isabellec}

\begin{figure}[t]
	\centerline{\includegraphics[width=3.2in]{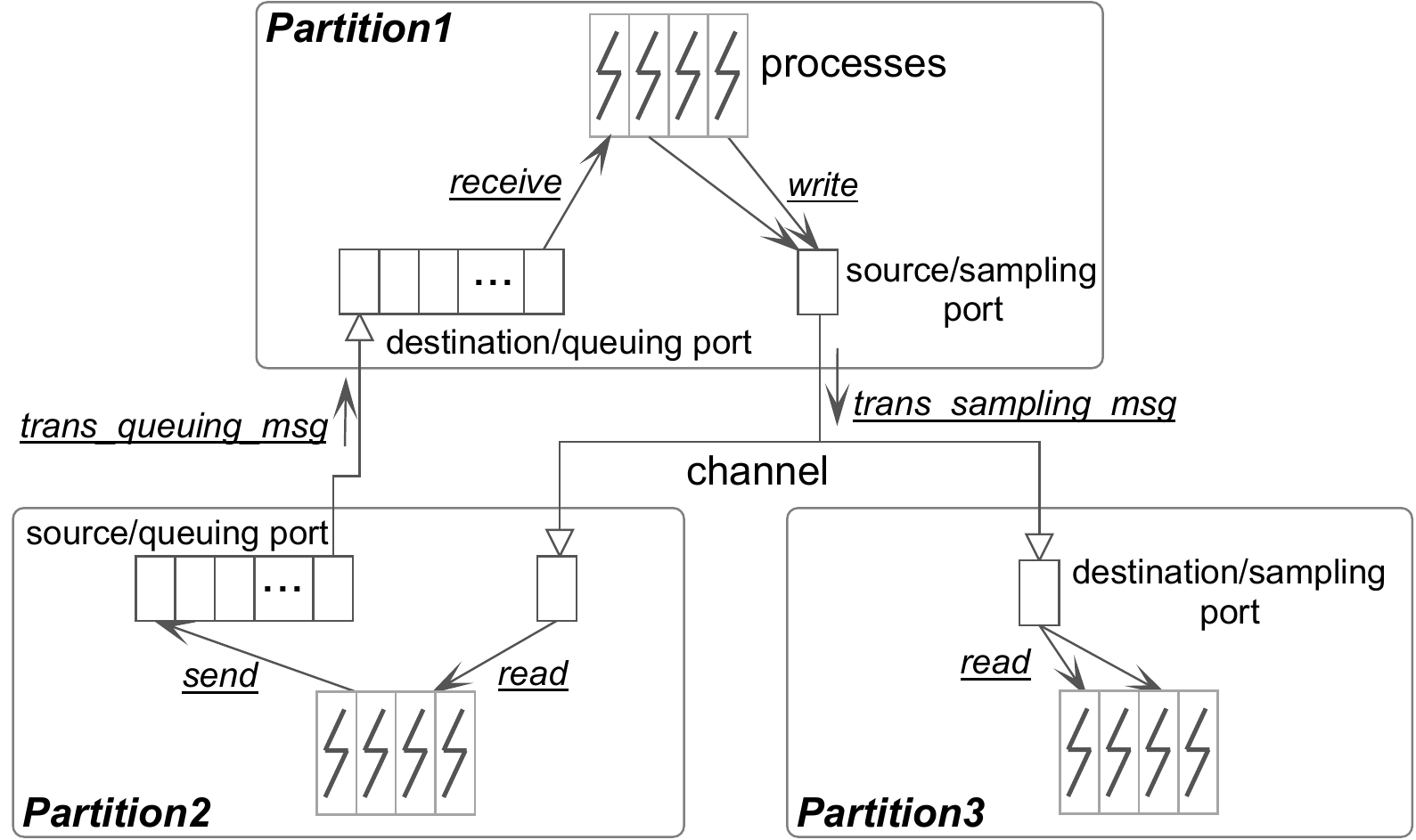}}
	\caption{Channel-based Communication in ARINC 653}
	\label{fig:comm}
\end{figure}

\textbf{Inter-partition Communication}:
The communication architecture is illustrated in {\figprefix} \ref{fig:comm}.
In the first stage of this work, we design the event specification completely based on the service behavior specified in ARINC 653. When proving the unwinding conditions on these events, we find covert channels ({\sectprefix} \ref{sect:reslt_disc} in detail).
We change the service specification in ARINC 653 by allowing message loss to avoid these covert channels according to the results in \cite{McCull88}. 

We use a set of functions to implement one service. For instance, the \emph{Send\_Queuing\_Message} service is implemented by the $send\_que\_msg\_lost$ function as follows. The manipulated port should be a source (\emph{is\_source\_port s p}) and queuing (\emph{is\_a\_queuingport s p}) port. It should belong to current partition too (\emph{is\_a\_port\_of\_partition s p}). If the message buffer of the port is full (\emph{is\_full\_portqueuing}), the service just omits the message. Otherwise, the message is inserted into the buffer (\emph{insert\_msg2queuing\_port}). 

\begin{isabellec}\isacodeftsz
\zipbeforecode
\isacommand{definition}\isamarkupfalse%
\ send\_que\_msg\_lost\ {\isacharcolon}{\isacharcolon}\ ``Sys\_Config\ {\isasymRightarrow}\ State\ {\isasymRightarrow}\ port\_id\ {\isasymRightarrow}\ Message\ {\isasymRightarrow}\ {\isacharparenleft}State\ {\isasymtimes}\ bool{\isacharparenright}''\ \isakeyword{where}\isanewline
\ \ \ \ {\isachardoublequoteopen}send\_que\_msg\_lost\ sc\ s\ p\ m\ {\isasymequiv}\ \isanewline
\ \ \ \ \ \ \ \ \ \ {\isacharparenleft}\textbf{if}{\isacharparenleft}{\isasymnot}\ is\_a\_queuingport\ s\ p\  {\isasymor}\ {\isasymnot}\ is\_source\_port\ s\ p \isanewline
\ \ \ \ \ \ \ \ \ \ \ \ \ \ \ \ {\isasymor}\ {\isasymnot}\ is\_a\_port\_of\_partition\ s\ p {\isacharparenright} \ \textbf{then} \ {\isacharparenleft}s{\isacharcomma}\ False{\isacharparenright}\isanewline
\ \ \ \ \ \ \ \ \ \ \ \textbf{else}\ \textbf{if}\ is\_full\_portqueuing\ sc\ s\ p\ \textbf{then} \ {\isacharparenleft}s{\isacharcomma}\ True{\isacharparenright}\isanewline
\ \ \ \ \ \ \ \ \ \ \ \textbf{else} \ {\isacharparenleft}insert\_msg2queuing\_port\ s\ p\ m{\isacharcomma}\ True{\isacharparenright}{\isacharparenright}{\isachardoublequoteclose}%
\zipaftercode
\end{isabellec}

The message transmission on channels is shown in {\figprefix} \ref{fig:comm}. 
For instance, the message transmission in queuing mode is specified as follows. If the source and destination port have been created (\emph{get\_portid\_by\_name\ s\ sn $\neq$ None}) and there are messages in the buffer of the source port (\emph{has\_msg\_inportqueuing}), a message in the buffer is removed (\emph{remove\_msg\_from\_queuingport}) and inserted into the buffer of the destination port. When the buffer of the destination port is full (\emph{is\_full\_portqueuing}), the message is discarded. 

\begin{isabellec}\isacodeftsz
\zipbeforecode
\isacommand{primrec}\isamarkupfalse%
\ transf\_que\_msg\_lost\ {\isacharcolon}{\isacharcolon}\ ``Sys\_Config\ {\isasymRightarrow}\ State\ {\isasymRightarrow}\ Channel\_Type \isanewline
{\isasymRightarrow}\ State''\ \isakeyword{where}\isanewline
\ \ {\isachardoublequoteopen}transf\_que\_msg\_lost\ sc\ s\ {\isacharparenleft}Queuing\ \_\ sn\ dn{\isacharparenright}\ {\isacharequal}\ \isanewline
\ \ \ \ {\isacharparenleft}\textbf{let}\ sp\ {\isacharequal}\ get\_portid\_by\_name\ s\ sn{\isacharsemicolon} dp\ {\isacharequal}\ get\_portid\_by\_name\ s\ dn\ \textbf{in}\isanewline
\ \ \ \ \ \ \textbf{if}\ sp\ {\isasymnoteq}\ None\ {\isasymand}\ dp\ {\isasymnoteq}\ None\ {\isasymand}\ has\_msg\_inportqueuing\ s\ {\isacharparenleft}the\ sp{\isacharparenright}\ \textbf{then}\isanewline
\ \ \ \ \ \ \ \ \textbf{let}\ sm\ {\isacharequal}\ remove\_msg\_from\_queuingport\ s\ {\isacharparenleft}the\ sp{\isacharparenright}\ \textbf{in}\isanewline
\ \ \ \ \ \ \ \ \ \ \ \ \textbf{if}\ is\_full\_portqueuing\ sc\ {\isacharparenleft}fst\ sm{\isacharparenright}\ {\isacharparenleft}the\ dp{\isacharparenright}\ \textbf{then} s \isanewline
\ \ \ \ \ \ \ \ \ \ \ \ \textbf{else}\ insert\_msg2queuing\_port\ {\isacharparenleft}fst\ sm{\isacharparenright}\ {\isacharparenleft}the\ dp{\isacharparenright}\ {\isacharparenleft}the\ {\isacharparenleft}snd\ sm{\isacharparenright}{\isacharparenright}\ \isanewline
\ \ \ \ \ \ \textbf{else}\ s {\isacharparenright}{\isachardoublequoteclose}\ {\isacharbar}\isanewline
\ \ {\isachardoublequoteopen}transf\_que\_msg\_lost\ sc\ s\ {\isacharparenleft}Sampling\ \_\ \_\ \_{\isacharparenright}\ {\isacharequal}\ s{\isachardoublequoteclose}
\zipaftercode
\vspace{-6pt}
\end{isabellec}

\subsection{Security Proofs}

Since the top-level specification is an instance of the security model, the first part of the security proofs is the instantiation proof. The assumptions 1) - 3) of the security model ({\defprefix} \ref{def:sec_model}) on the interference relation are preserved by the $interference1$ function. The assumption 4) is preserved by $\sim$ for the scheduler. The definition of $\sim$ is an equivalence relation at the top level, which means the preservation of the assumption 5). By the following lemma, the assumption 6) of the security model is preserved by the top-level specification.

\begin{lemma}[Event Enabled in Top-level Specification]
{
\begin{equation*}
\forall s \ e . \ \reachable(s) \longrightarrow (\exists s'. \ (s, s') \in exec\_event(e))
\end{equation*}
}
\end{lemma}

The second step of the security proofs is to show the UCEs, i.e. satisfaction of {\defprefix}s \ref{def:sc_e} and \ref{def:lr_e}.
We define a set of \emph{concrete conditions} for events. Satisfaction of the concrete conditions of one event implies that the event satisfies the UCEs. For instance, {\defprefix} \ref{def:crtqueport_sc} shows the concrete condition of step consistent for the \emph{Create\_Queuing\_Port} event, which is an instance of UCEs of the event.  

\begin{definition}[Concrete $SC(e)$ of Creating\_Queuing\_Port]
\label{def:crtqueport_sc}
\end{definition}
{
\small
\begin{equation*}
\begin{split}
{\isasymforall}\ d\ & s\ t \ s' \ t' \ p.\ \reachable(s) \wedge \reachable(t) \wedge \equidom{s}{d}{t} \wedge \equidom{s}{\sched}{t} \\
& \wedge is\_part\ conf\ (cur \ s) \wedge (cur \ s) \interf d \wedge
\equidom{s}{cur \ s}{t} \\
& \wedge s' = fst\ (create\_queuing\_port\ \sysconf\ s\ p) \\
& \wedge t' = fst\ (create\_queuing\_port\ \sysconf\ t\ p) \\
& \longrightarrow \equidom{s'}{d}{t'}
\end{split}
\end{equation*}
}

Finally, we conclude the satisfaction of the \emph{noninfluence} property on the top-level specification and all other information-flow security properties according to the inference framework of the security model. 

\section{Second-level Specification and Security Proofs}
\label{sect:2ndspec}
In the second-level specification, we refine the top-level one by adding processes, process scheduling in partitions, and process management services in ARINC 653. We first instantiate the security model as the kernel execution model. Then, we present the refinement. Finally, the security proofs of the specification are discussed.

\subsection{Kernel Execution Model}
In ARINC 653, a partition comprises one or more processes that combine dynamically to provide the functions associated with that partition \cite{ARINC653p1}. Processes are separated among partitions, while processes within a partition share the resources of the partition. Process management provides the services to control the life-cycle of processes. We refine the state at the top-level as follows. The state variables at the top level are not changed. The state is extended by new state variables of processes by means of \textbf{record} extension. Thus, the state relation $R$ is simple and defined as follows.
A partition has a set of created processes ($procs$) and may have a current executing process ($cur\_proc\_part$). A process has a state ($proc\_state$). We found a covert channel if we use global process identifiers ({\sectprefix} \ref{sect:reslt_disc} in detail). Here, we use separated process identifiers for each partition. 

\begin{isabellec}
\isacodeftsz
\zipbeforecode
\isacommand{record}\isamarkupfalse%
\ StateR\ {\isacharequal}\ State\ {\isacharplus}\isanewline
 \ \ \ \ \ \ \ \ \ \ \ \ procs\ {\isacharcolon}{\isacharcolon}\ {\isachardoublequoteopen}partition{\isacharunderscore}id\ {\isasymrightharpoonup}\ {\isacharparenleft}process{\isacharunderscore}id\ set{\isacharparenright}{\isachardoublequoteclose}\isanewline
 \ \ \ \ \ \ \ \ \ \ \ \ cur{\isacharunderscore}proc{\isacharunderscore}part\ {\isacharcolon}{\isacharcolon}\ {\isachardoublequoteopen}partition{\isacharunderscore}id\ {\isasymrightharpoonup}\ process{\isacharunderscore}id{\isachardoublequoteclose}\isanewline
 \ \ \ \ \ \ \ \ \ \ \ \ proc{\isacharunderscore}state\ {\isacharcolon}{\isacharcolon}\ {\isachardoublequoteopen}partition{\isacharunderscore}id\ {\isasymtimes}\ process{\isacharunderscore}id\ {\isasymrightharpoonup}\ Proc{\isacharunderscore}State{\isachardoublequoteclose}
\zipaftercode
\end{isabellec}

\begin{isabellec}
\isacodeftsz
\zipbeforecode
\isacommand{definition}\isamarkupfalse%
\ R\ {\isacharcolon}{\isacharcolon}\ {\isachardoublequoteopen}StateR\ {\isasymRightarrow}\ State{\isachardoublequoteclose}\ {\isacharparenleft}{\isachardoublequoteopen}{\isasymUp}{\isacharunderscore}{\isachardoublequoteclose}\ {\isacharbrackleft}{\isadigit{5}}{\isadigit{0}}{\isacharbrackright}{\isacharparenright}\ \isakeyword{where} \isanewline
\ \ \ {\isachardoublequoteopen}R\ r\ {\isacharequal}\ {\isasymlparr}cur\ {\isacharequal}\ cur\ r{\isacharcomma}\ partitions\ {\isacharequal}\ partitions\ r{\isacharcomma}\isanewline
\ \ \ \ \ \ \ \ \ \ \ \ \ comm\ {\isacharequal}\ comm\ r{\isacharcomma}\ part{\isacharunderscore}ports\ {\isacharequal}\ part{\isacharunderscore}ports\ r {\isasymrparr}{\isachardoublequoteclose}
\zipaftercode
\end{isabellec}

Each event at the top level is refined by one at the second level. We add events for process scheduling and process management services. 
The $exec\_event$ function at the top level is extended by adding maps of new events to new functions in the event specification. The event domain and the interference relation at the second level are the same as those at the top level. The state equivalence relation is extended on new state variables as $\equidomR{s}{d}{t}$ as follows. The state equivalence on new state variables, i.e. $\equidomsub{s}{d}{\Delta}{t}$ in {\defprefix} \ref{def:refine}, is defined as $\equidomRsub{s}{d}{\Delta}{t}$ at the second level. It requires that the processes of a partition $d$, the states of processes, and the current executing process in the partition are the same in states $s$ and $t$.  

\begin{isabellec}
\isacodeftsz
\zipbeforecode
\isacommand{definition}\isamarkupfalse%
\ vpeqR\ {\isacharcolon}{\isacharcolon}\ {\isachardoublequoteopen}StateR\ {\isasymRightarrow}\ domain{\isacharunderscore}id\ {\isasymRightarrow}\ StateR\ {\isasymRightarrow}\ bool{\isachardoublequoteclose}\ {\isacharparenleft}{\isachardoublequoteopen}{\isacharparenleft}{\isacharunderscore}\ {\isasymsim}{\isachardot}\ {\isacharunderscore}\ {\isachardot}{\isasymsim}\ {\isacharunderscore}{\isacharparenright}{\isachardoublequoteclose}{\isacharparenright}\isanewline
\ \ \isakeyword{where}\ {\isachardoublequoteopen}vpeqR\ s\ d\ t\ {\isasymequiv}\ \ {\isacharparenleft}R \ s{\isacharparenright}\ {\isasymsim} \ d\ {\isasymsim} {\isacharparenleft}R \ t{\isacharparenright}\ {\isasymand}\ {\isacharparenleft}s{\isasymsim}{\isachardot}d{\isachardot}{\isasymsim}\isactrlsub {\isasymDelta}t{\isacharparenright}{\isachardoublequoteclose}
\zipaftercode
\vspace{-6pt}
\end{isabellec}

\subsection{Event Specification}
\begin{figure}[t]
\centerline{\includegraphics[width=3.4in]{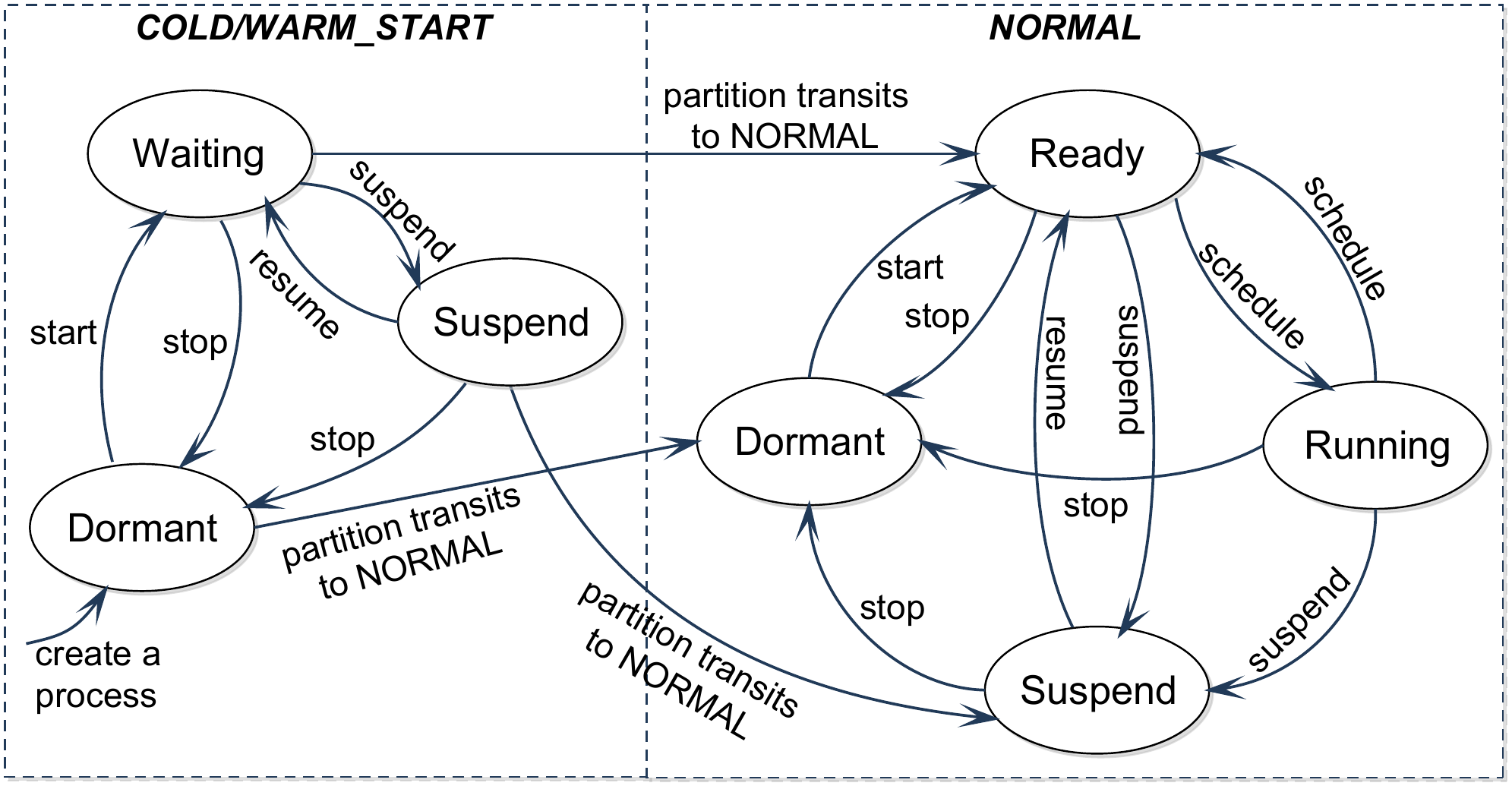}}
\caption{Process State Transitions in Second-level Specification}
\label{fig:proc_states}
\end{figure}

Due to the new events introduced in the refinement, we use new functions to implement the process management and scheduling. The existing functions in the top-level specification are refined according to the new state type. The process state transitions in the second-level specification are shown in {\figprefix} \ref{fig:proc_states}. We capture all state transitions defined in ARINC 653 except those triggered by intra-partition communication, which will be modeled in future refinement. 

\textbf{Process Management}:
Seven basic services of process management in ARINC 653, i.e. \emph{Create/\allowbreak Start/\allowbreak Stop/\allowbreak Suspend/\allowbreak Resume\_Process}, \emph{Set\_Priority}, and \emph{Get\_Process\_Status}, are modeled at the second level. We use a set of functions to implement these services. For instance, \emph{Start\_Process} is implemented by the $start\_process$ function as follows. 

\begin{isabellec}
\isacodeftsz
\zipbeforecode
\isacommand{definition}\isamarkupfalse%
\ start{\isacharunderscore}process\ {\isacharcolon}{\isacharcolon}\ {\isachardoublequoteopen}StateR\ {\isasymRightarrow}\ process{\isacharunderscore}id\ {\isasymRightarrow}\ StateR{\isachardoublequoteclose}\ \isakeyword{where}\isanewline
\ \ {\isachardoublequoteopen}start{\isacharunderscore}process\ s\ p\ {\isasymequiv}\isanewline
\ \ \ \ \ \textbf{if}\ p\ {\isasymin}\ the\ {\isacharparenleft}{\isacharparenleft}procs\ s{\isacharparenright}\ {\isacharparenleft}cur\ s{\isacharparenright}{\isacharparenright} {\isasymand}\isanewline
\ \ \ \ \ \ \ \ {\isacharparenleft}state\ {\isacharparenleft}the\ {\isacharparenleft}{\isacharparenleft}proc{\isacharunderscore}state\ s{\isacharparenright}\ {\isacharparenleft}cur\ s{\isacharcomma}\ p{\isacharparenright}{\isacharparenright}{\isacharparenright}{\isacharparenright}\ {\isacharequal}\ DORMANT\ \textbf{then}\isanewline
\ \ \ \ \ \ \ \ \ \ \ \textbf{let}\ st\ {\isacharequal}\ {\isacharparenleft}if\ the\ {\isacharparenleft}{\isacharparenleft}partitions\ s{\isacharparenright}\ {\isacharparenleft}cur\ s{\isacharparenright}{\isacharparenright}\ {\isacharequal}\ NORMAL\isanewline
\ \ \ \ \ \ \ \ \ \ \ \ \ \ \ \ \ \ \ \ \ \ \ \ then\ READY\ else\ WAITING{\isacharparenright}{\isacharsemicolon}\isanewline
\ \ \ \ \ \ \ \ \ \ \ \ \ \ \ \ pst\ {\isacharequal}\ the\ {\isacharparenleft}{\isacharparenleft}proc{\isacharunderscore}state\ s{\isacharparenright}\ {\isacharparenleft}cur\ s{\isacharcomma}\ p{\isacharparenright}{\isacharparenright}{\isacharsemicolon}\isanewline
\ \ \ \ \ \ \ \ \ \ \ \ \ \ \ \ proc{\isacharunderscore}state{\isacharprime}\ {\isacharequal}\ {\isacharparenleft}proc{\isacharunderscore}state\ s{\isacharparenright}\ {\isacharparenleft}{\isacharparenleft}cur\ s{\isacharcomma}\ p{\isacharparenright}\ {\isacharcolon}{\isacharequal}\isanewline
\ \ \ \ \ \ \ \ \ \ \ \ \ \ \ \ \ \ \ \ \ \ \ \ \ \ \ \ \ \ \ \ \ \ \ \ \ \ \ \ \ \ \ \ \ \ \ \ \ \ \ \ \ \ \ \ \ \ \ \ \ Some\ (pst \ {\isasymlparr}state\ :{\isacharequal}\ st{\isasymrparr}{\isacharparenright}{\isacharparenright}\ \textbf{in}\isanewline
\ \ \ \ \ \ \ s{\isasymlparr}proc{\isacharunderscore}state{\isacharcolon}{\isacharequal}proc{\isacharunderscore}state{\isacharprime}{\isasymrparr} \ \textbf{else}\ s{\isachardoublequoteclose}
\zipaftercode
\end{isabellec}

The process to be started should belong to the current executing partition and be in \emph{DORMANT} state. As shown in {\figprefix} \ref{fig:proc_states}, if the current partition is in \emph{NORMAL} mode, the new state of the process is \emph{READY}. Otherwise, the new state is \emph{WAITING}.  

\textbf{Process Scheduling}:
ARINC 653 defines a priority based process scheduling in partitions, which is implemented as follows. The process scheduling occurs in a partition only when the partition is in \emph{NORMAL} mode. It first sets the state of the current executing process to \emph{READY} (\emph{setRun2Ready}). Then it chooses one of processes in \emph{READY} state with highest priority in the current partition. Finally, it sets the chosen process as the current executing process of the current partition.

\begin{isabellec}
\isacodeftsz
\zipbeforecode
\isacommand{definition}\isamarkupfalse%
\ schedule{\isacharunderscore}process\ {\isacharcolon}{\isacharcolon}\ {\isachardoublequoteopen}StateR\ {\isasymRightarrow}\ StateR\ set{\isachardoublequoteclose}\ \isakeyword{where}\isanewline
\ \ {\isachardoublequoteopen}schedule{\isacharunderscore}process\ s\ {\isasymequiv}\isanewline
\ \ \ \ \textbf{if}\ is{\isacharunderscore}part\ conf\ {\isacharparenleft}cur\ s{\isacharparenright}\  {\isasymand} \isanewline
\ \ \ \ \ \ \ \ \ {\isacharparenleft}the\ {\isacharparenleft}{\isacharparenleft}partitions\ s{\isacharparenright}\ {\isacharparenleft}cur\ s{\isacharparenright}{\isacharparenright}{\isacharparenright}\ {\isacharequal}\ NORMAL\ \textbf{then}\isanewline
\ \ \ \ \ \ {\isacharparenleft}\textbf{let}\ s{\isacharprime}\ {\isacharequal}\ setRun{\isadigit{2}}Ready\ s{\isacharsemicolon}\isanewline
\ \ \ \ \ \ \ \ \ \ readyprs\ {\isacharequal}\ {\isacharbraceleft}p{\isachardot}\ p{\isasymin}the\ {\isacharparenleft}procs\ s{\isacharprime}\ {\isacharparenleft}cur\ s{\isacharprime}{\isacharparenright}{\isacharparenright}\ {\isasymand}\ \isanewline
\ \ \ \ \ \ \ \ \ \ \ \ \ \ \ \ \ \ \ \ \ \ \ \ \  state\ {\isacharparenleft}the\ {\isacharparenleft}proc{\isacharunderscore}state\ s{\isacharprime}\ {\isacharparenleft}cur\ s{\isacharprime}{\isacharcomma}p{\isacharparenright}{\isacharparenright}{\isacharparenright}\ {\isacharequal}\ READY{\isacharbraceright}{\isacharsemicolon}\isanewline
\ \ \ \ \ \ \ \ \ \ selp\ {\isacharequal}\ SOME\ p{\isachardot}\isanewline
\ \ \ \ \ \ \ \ \ \ \ \ \ \ \ p{\isasymin}{\isacharbraceleft}x{\isachardot}\ state\ {\isacharparenleft}the\ {\isacharparenleft}proc{\isacharunderscore}state\ s{\isacharprime}\ {\isacharparenleft}cur\ s{\isacharprime}{\isacharcomma}x{\isacharparenright}{\isacharparenright}{\isacharparenright}\ {\isacharequal}\ READY\ {\isasymand}\isanewline
\ \ \ \ \ \ \ \ \ \ \ \ \ \ \ \ \ \ {\isacharparenleft}{\isasymforall}y{\isasymin}readyprs{\isachardot}\ priority\ {\isacharparenleft}the\ {\isacharparenleft}proc{\isacharunderscore}state\ s{\isacharprime}\ {\isacharparenleft}cur\ s{\isacharprime}{\isacharcomma}x{\isacharparenright}{\isacharparenright}{\isacharparenright}\ {\isasymge}\isanewline
\ \ \ \ \ \ \ \ \ \ \ \ \ \ \ \ \ \ \ \ \ \ \ \ \ \ \ \ \ \ \ \ \ \ \ \ \ priority\ {\isacharparenleft}the\ {\isacharparenleft}proc{\isacharunderscore}state\ s{\isacharprime}\ {\isacharparenleft}cur\ s{\isacharprime}{\isacharcomma}y{\isacharparenright}{\isacharparenright}{\isacharparenright}{\isacharparenright}{\isacharbraceright}{\isacharsemicolon}\isanewline
\ \ \ \ \ \ \ \ \ \ st\ {\isacharequal}\ the\ {\isacharparenleft}{\isacharparenleft}proc{\isacharunderscore}state\ s{\isacharprime}{\isacharparenright}\ {\isacharparenleft}cur\ s{\isacharprime}{\isacharcomma}\ selp{\isacharparenright}{\isacharparenright}{\isacharsemicolon}\isanewline
\ \ \ \ \ \ \ \ \ \ proc{\isacharunderscore}st\ {\isacharequal}\ proc{\isacharunderscore}state\ s{\isacharprime}{\isacharsemicolon}
 \ cur{\isacharunderscore}pr\ {\isacharequal}\ cur{\isacharunderscore}proc{\isacharunderscore}part\ s{\isacharprime}\ \textbf{in}\isanewline
\ \ \ \ \ \ {\isacharbraceleft}s{\isacharprime}{\isasymlparr}proc{\isacharunderscore}state\ {\isacharcolon}{\isacharequal}\isanewline
\ \ \ \ \ \ \ \ \ \ \ \ \ \ \ proc{\isacharunderscore}st\ {\isacharparenleft}{\isacharparenleft}cur\ s{\isacharprime}{\isacharcomma}\ selp{\isacharparenright}\ {\isacharcolon}{\isacharequal}\ Some\ {\isacharparenleft}st{\isasymlparr}state\ {\isacharcolon}{\isacharequal}\ RUNNING{\isasymrparr}{\isacharparenright}{\isacharparenright}{\isacharcomma}\isanewline
\ \ \ \ \ \ \ \ \ \ \ cur{\isacharunderscore}proc{\isacharunderscore}part\ {\isacharcolon}{\isacharequal}\ cur{\isacharunderscore}pr{\isacharparenleft}cur\ s{\isacharprime}\ {\isacharcolon}{\isacharequal}\ Some\ selp{\isacharparenright}{\isasymrparr}{\isacharbraceright}{\isacharparenright}\isanewline
\ \ \ \ \textbf{else} \ {\isacharbraceleft}s{\isacharbraceright}{\isachardoublequoteclose}
\zipaftercode
\end{isabellec}

\textbf{Refined Events}:
All events at the top level are refined according to the new state type. For instance, the $set\_partition\_mode$ function ({\subsectprefix} \ref{subsect:evt_spc_top}) at the top level is refined as follows. When setting a partition to \emph{NORMAL} mode, the states of processes in the partition are correctly set (\emph{set\_procs\_to\_normal}) according to process state transitions in {\figprefix} \ref{fig:proc_states}. When setting a partition from \emph{NORMAL} mode to others, all processes in the partition are deleted (\emph{remove\_partition\_resources}). 

\begin{isabellec}
\isacodeftsz
\zipbeforecode
\isacommand{definition}\isamarkupfalse%
\ set{\isacharunderscore}partition{\isacharunderscore}modeR\ {\isacharcolon}{\isacharcolon}\ {\isachardoublequoteopen}Sys{\isacharunderscore}Config\ {\isasymRightarrow}\ StateR\ {\isasymRightarrow}\ partition{\isacharunderscore}mode{\isacharunderscore}type\ {\isasymRightarrow}\ StateR{\isachardoublequoteclose}\ \isakeyword{where}\isanewline
\ \ {\isachardoublequoteopen}set{\isacharunderscore}partition{\isacharunderscore}modeR\ sc\ s\ m\ {\isasymequiv}\ \isanewline
\ \ \ \ \ \ {\isacharparenleft}\textbf{if}\ {\isacharparenleft}partitions\ s{\isacharparenright}\ {\isacharparenleft}cur\ s{\isacharparenright}\ {\isasymnoteq}\ None\ {\isasymand}\isanewline
\ \ \ \ \ \ \ \ \ \ {\isasymnot}\ {\isacharparenleft}the\ {\isacharparenleft}{\isacharparenleft}partitions\ s{\isacharparenright}\ {\isacharparenleft}cur\ s{\isacharparenright}{\isacharparenright}\ {\isacharequal}\ COLD{\isacharunderscore}START \isanewline
\ \ \ \ \ \ \ \ \ \ \ \ \ \ \ \ {\isasymand}\ m\ {\isacharequal}\ WARM{\isacharunderscore}START{\isacharparenright}\ \textbf{then}\isanewline
\ \ \ \ \ \ \ \ \textbf{let}\ pts\ {\isacharequal}\ partitions\ s{\isacharsemicolon}\isanewline
\ \ \ \ \ \ \ \ \ \ \ \ s{\isacharprime}\ {\isacharequal}\ {\isacharparenleft}\textbf{if}\ m\ {\isacharequal}\ NORMAL\ \textbf{then}\ \isanewline
\ \ \ \ \ \ \ \ \ \ \ \ \ \ \ \ \ \ \ \ \ \ set{\isacharunderscore}procs{\isacharunderscore}to{\isacharunderscore}normal\ s\ {\isacharparenleft}cur\ s{\isacharparenright}\isanewline
\ \ \ \ \ \ \ \ \ \ \ \ \ \ \ \ \ \ \ \textbf{else}\ \textbf{if}\ the\ {\isacharparenleft}{\isacharparenleft}partitions\ s{\isacharparenright}\ {\isacharparenleft}cur\ s{\isacharparenright}{\isacharparenright}\ {\isacharequal}\ NORMAL\ \textbf{then}\isanewline
\ \ \ \ \ \ \ \ \ \ \ \ \ \ \ \ \ \ \ \ \ \ remove{\isacharunderscore}partition{\isacharunderscore}resources\ s\ {\isacharparenleft}cur\ s{\isacharparenright}\ \isanewline
\ \ \ \ \ \ \ \ \ \ \ \ \ \ \ \ \ \ \ \textbf{else}\ s\ {\isacharparenright} \ \textbf{in} \isanewline
\ \ \ \ \ \ \ \ \ s{\isacharprime}{\isasymlparr}partitions\ {\isacharcolon}{\isacharequal}\ pts{\isacharparenleft}cur\ s{\isacharprime}\ {\isacharcolon}{\isacharequal}\ Some \ m {\isacharparenright}{\isasymrparr}\isanewline
\ \ \ \ \ \ \ \textbf{else} \ s{\isacharparenright}{\isachardoublequoteclose}
\zipaftercode
\end{isabellec}

\subsection{Security Proofs}

Since the second-level specification is a refinement and an instance of the security model, the first part of the security proofs are the instantiation and refinement proofs. The assumptions (1) - (6) of the security model ({\defprefix} \ref{def:sec_model}) have been proven on the second-level specification. In order to show the refinement relation, conditions (1) - (6) in {\defprefix} \ref{def:refine} are proven on the second-level specification. 

The second step of security proofs is to show the UCEs, i.e. satisfaction of {\defprefix}s \ref{def:sc_e} and \ref{def:lr_e}. By following {\theoremprefix} \ref{thm:sec_conc_spec}, we only need to prove that $SK_C$ satisfies $SC_{\Delta} \wedge LR_{\Delta}$ to show the information-flow security, since the satisfaction of $SC_A \wedge LR_A$ in $SK_A$ has been proven at the top level and we have $SK_A \refine SK_C$. 
Therefore, we define a set of \emph{concrete conditions} for all events on the new state variables. Satisfaction of the conditions of one event implies that the event satisfies the unwinding conditions. For instance, {\defprefix} \ref{def:crtqueport_scdlt} shows the concrete condition of step consistent for the \emph{Schedule\_Process} event, which is an instance of UCEs on the new state variables ($SC_{C\Delta}(e)$) of the event.  

\begin{definition}[Concrete $SC_{C\Delta}(e)$ of Schedule\_Process]
\label{def:crtqueport_scdlt}
\end{definition}
{
\small
\begin{equation*}
\begin{split}
\forall \ d\ & s\ t \ s' \ t' \ p. \ \reachable_C(s) \wedge \reachable_C(t) \wedge \equidomR{s}{d}{t} \wedge \equidomR{s}{\sched}{t} \\
& \wedge is\_part\ conf\ (cur \ s) \wedge (cur \ s) \interf d \wedge
\equidomR{s}{cur \ s}{t} \\
& \wedge s' \in schedule\_process \ s \wedge t' \in schedule\_process \ t \\
& \longrightarrow \equidomRsub{s'}{d}{\Delta}{t'}
\end{split}
\end{equation*}
}

Finally, we conclude the satisfaction of the \emph{noninfluence} property on the second-level specification and all other information-flow security properties according to {\theoremprefix} \ref{thm:sec_conc_spec} and the inference framework of the security model. 

\section{Results and Discussion}
\label{sect:reslt_disc}

\subsection{Evaluation}
We use Isabelle/HOL as the specification and verification system for {\skname}s. 
The proofs are conducted in the structured proof language \emph{Isar} in Isabelle, allowing for proof text naturally understandable for both humans and computers. All derivations of our proofs have passed through the Isabelle proof kernel. 

\begin{table}[t]
\centering
\footnotesize
\caption{Specification and Proof Statistics} 
\begin{tabular} {|c|c|c|c|c|c|}
\hline
\multirow{2}{*}{\textbf{Item}} & \multicolumn{2}{c|}{\textbf{Specification}} & \multicolumn{2}{c|}{\textbf{Proof}} & \multirow{2}{*}{\textbf{PM}}
\\ \cline{2-5} 
 & \tabincell{c}{\bfseries \# of type/ \\ \bfseries definition} & \bfseries LOC & \tabincell{c}{\bfseries \# of lemma \\  \bfseries/theorem} & \bfseries LOP & 
\\ \hline 
\tabincell{c}{Security \\ Model}  & 25 & 130 & 63 & 900 & 1 
\\\hline
\tabincell{c}{Top-level \\ Specification} & 116 & 680 & 193 & 5,200 & 6
\\\hline
Refinement & 6 & 100 & 42 & 600 & 1
\\\hline
\tabincell{c}{2nd-level \\ Specification} & 45 & 330 & 111 & 1,100 & 4
\\\hline
\textbf{Total} & 192 & 1,240 & 409 & 7,800 & 12
\\\hline
\end{tabular}
\label{tbl:stat}
\end{table}

\begin{figure}[t]
\centerline{\includegraphics[width=3.2in]{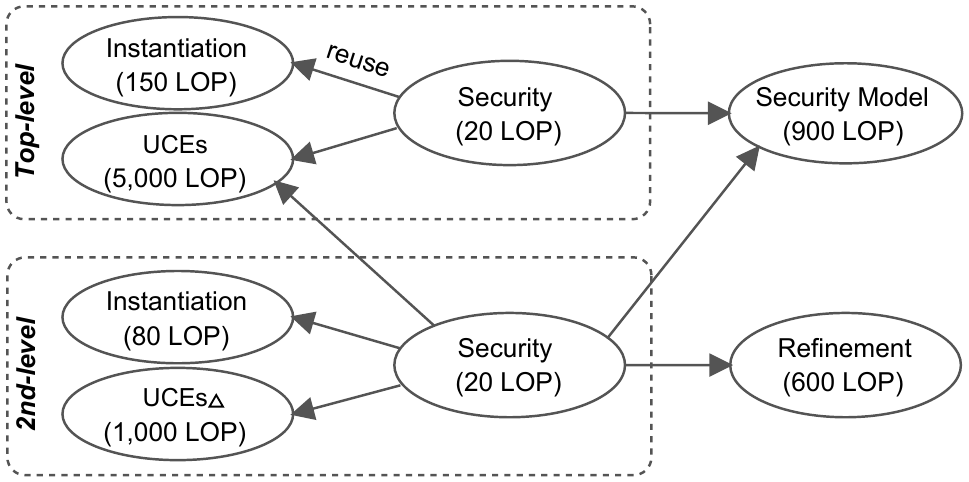}}
\caption{Proof Reusability}
\label{fig:proof}
\end{figure}

The statistics for the effort and size of the specification and proofs are shown in {\tableprefix} \ref{tbl:stat}.
We use 190 datatypes/definitions and $\sim$ 1,240 lines of code (LOC) of Isabelle/HOL to develop the security model, the refinement framework, and two levels of functional specifications. 409 lemmas/theorems in Isabelle/HOL are proven using $\sim$ 7,800 lines of proof (LOP) of Isar to ensure the information-flow security of the specification. 
The work is carried out by a total effort of roughly 12 person-months (PM). 
The proof reusability is shown in {\figprefix} \ref{fig:proof}. The proof of the security model is reusable at each level. The proof of our refinement is reusable at each refinement step. The major part of proof at each level is UCEs which is reusable at a lower level, for example $\sim$ 5,000 LOP of the top-level specification is reused in the second-level specification.

\subsection{Covert Channels and Attack Analysis}
\label{subsec:covt_ch_arinc}

When proving the UCEs in our original specification which is completely compliant with ARINC 653, a few covert channels are found in the standard. Then, we conduct a code-to-spec review on VxWorks 653, XtratuM, and POK in accordance with our formal specification. The covert channels are also found in them. {\tableprefix} \ref{tbl:covertchannels} shows the covert channels and their existence. 

\begin{table}[t]
\centering
\footnotesize
\caption{The Found Covert Channels} 
\begin{tabular} {|c|c|c|c|c|}
\hline
\tabincell{c}{\bfseries Covert \\ \bfseries Channel}  & \bfseries ARINC standard & \bfseries VxWorks 653 & \bfseries XtratuM & \bfseries POK 
\\\hline 
(1) & $\exist$ & $\exist$ & $\exist$ & $\exist$ 
\\\hline
(2) & $\exist$ & $\exist$ &  &  
\\\hline
(3) & $\halfexist$ & $\exist$ &  &  
\\\hline
(4) & & $\halfexist$ & $\halfexist$ & 
\\\hline
(5) & $\halfexist$ & & & 
\\\hline
(6) & $\exist$ & $\exist$ & &  
\\\hline
\multicolumn{5}{l}{
$\exist$: existing; $\halfexist$: potential; \emph{blank}: not existing
}
\end{tabular}
\label{tbl:covertchannels}
\end{table} 

Covert channels 1, 2 and 6 actually exist in the ARINC 653 standard. Covert channels 3 and 5 are potential ones and may be introduced to the specification by careless design. Covert channel 4 does not exist in ARINC 653. However, it is a potential flaw that should be taken into account. They are described as follows. In Appendix A, we present how we find and fix them in our specification in detail. 

\begin{enumerate}

\item \textbf{Queuing mode channel}: 
If there is a queuing mode channel from a partition $a$ to a partition $b$, we find a covert channel from $b$ to $a$ in ARINC 653. 

\item \textbf{Leakage of port identifiers}:
In the \emph{Send/Receive\_Queuing\_Message} and \emph{Write/Read\_Sampling\_Message} services, there is no judgement on whether the accessing port belongs to the current partition. The port identifier in ARINC 653 is a covert channel. 

\item \textbf{Shared space of port identifiers}:
In ARINC 653, the \emph{Create\_Sampling\_Port} and \emph{Create\_Queuing\_Port} services create a port and return a new unique identifier assigned by the kernel to the new port. A potential covert channel is to use a global variable for unused port identifiers.

\item \textbf{Partition scheduling}:
State related scheduling policies, such as policies only selecting non-\emph{IDLE} partitions, is a covert channel.

\item \textbf{Shared space of process identifiers}:
In ARINC 653, the \emph{Create\_Process} service creates a process and assigns an unique identifier to the created process. 
A potential covert channel is to use a global variable for unused process identifiers.

\item \textbf{Leakage of process identifiers}:
In the services of process management, there is no judgement on whether the process belongs to the current partition. 
Thus, the \emph{locally respects} condition is not preserved on the events. The process identifier in ARINC 653 becomes a covert channel.

\end{enumerate}

Then we review the source code of implementations to validate the covert channels. As we consider single-core separation kernels in this paper, we manually review the single-core version of the implementations. Since the reviewed implementations are non-preemptive during the execution of hypercalls and have the same execution model as in our specification, it makes sense that we review the implementations according to our specification. 
In EAL 7 evaluation of {\skname}s, to ensure that the proved security really means that the implementation has the appropriate behavior, a formal model of the low-level design is created. Then, the correspondence between the model and the implementation is shown. Since our specification is at high level, we use the unwinding conditions to manually check the source code of hypercalls in the implementations, rather than to show their correspondence.

The result of code review is shown in {\tableprefix} \ref{tbl:covertchannels}.
The version of VxWorks 653 Platform we review is v2.2. The covert channel 5 is not found during code review. However, the covert channel 4 potentially exists and the other four covert channels are found in VxWorks 653. 
The version of XtratuM we review is v3.7.3. The covert channel 1 exists and the covert channel 4 is a potential one in XtratuM. 
The version of POK we review is the latest one released in 2014. We find the covert channel 1 in POK.
In Appendix B, we present where the covert channels exist in the source code and how we find them in detail.

The found covert channels pose threats to real-world {\skname}s. 
In order to analyze the potential attacks, we assume a threat model of {\skname}s in which all user-level code in partitions is malicious and acting to break the security policy. The attacker's goal is to read or infer the information in partitions that should remain secret to it according to the policy. 
Malicious programs in partitions could utilise the covert channels to break the security policy. 
The security risk of covert channels 1, 2, and 6 is high. Security information in a secret partition can be easily leaked by attackers. The covert channel 4 is a timing channel. Covert channels 3 and 5 have low bandwidth in real-world systems. 
We illustrate potential attacks to {\skname}s in Appendix C in detail.

\subsection{Discussion}

The refinement-based specification and analysis method in the paper is compliant to the EAL 7 of CC certification. 
With regard to the EAL 7, the main requirements addressed by formal methods are (1) a formal security policy model (SPM) of Target of Evaluation (TOE), (2) a complete semi-formal functional specification (FSP) with an additional formal specification, and (3) a complete semi-formal and modular design with high-level TOE design specification (TDS). 
The security model in this paper represents the security policies of {\skname}s and is a formal model of the SPM. The two levels of functional specifications in this paper correspond to the FSP. 
The properties demonstrated on the SPM are formally preserved down to the FSP by the model instantiation and instantiation proofs. 
The functional specification can be refined to a TDS model using the stepwise refinement. The refinement provides the formal link between the FSP and the TDS. Finally, code-to-spec review can be considered between the last formal model of the TDS and the implementation to show the correspondence. In this paper, we conduct a code review of the implementations according to our specification.

The method in this paper can alleviate the efforts of CC certification. 
The instantiation of the security model and the refinement framework ease the development and make the proof easier. 
As the same in the high assurance levels of CC certification of INTEGRITY-178B and AAMP7G \cite{Rich04}, formal model and proof are part of the evaluation and directly submitted to the evaluation team for certification. 
Certainly, formal model and proof should be created in accordance with a set of ``safe'' rules, such as rules of Isabelle/HOL in \cite{Blas15}, which have been complied with by our specification. On the other hand, for a specific TOE in CC certification, our specification may be revised and TDS model has to be developed.

\section{Conclusions and Future Work}
\label{sect:concl}
In this paper, we have presented a refinement-based specification development and security analysis method for {\skname}s. By the proposed superposition refinement, we have developed two levels of formal specification. 
We provided the mechanically checked proofs of information-flow security to overcome covert channels in separation kernels. We revealed some security flaws in ARINC 653 and implementations. 
In the next step, we will refine the second-level specification to lower levels by adding services of complicated process management and intra-partition communication. Supporting multi-core is also under consideration in future. 
Due to the kernel concurrency between cores, we are developing rely-guarantee based approach to specify and verify multi-core separation kernels.

\section*{Acknowledgement}
We would like to thank David Basin of Department of Computer Science, ETH Zurich, Gerwin Klein and Ralf Huuck of NICTA, Australia for their suggestions. 

\ifCLASSOPTIONcaptionsoff
  \newpage
\fi


\vspace*{-4\baselineskip}
\begin{IEEEbiography}[{\includegraphics[width=1in,height=1.25in,clip,keepaspectratio]{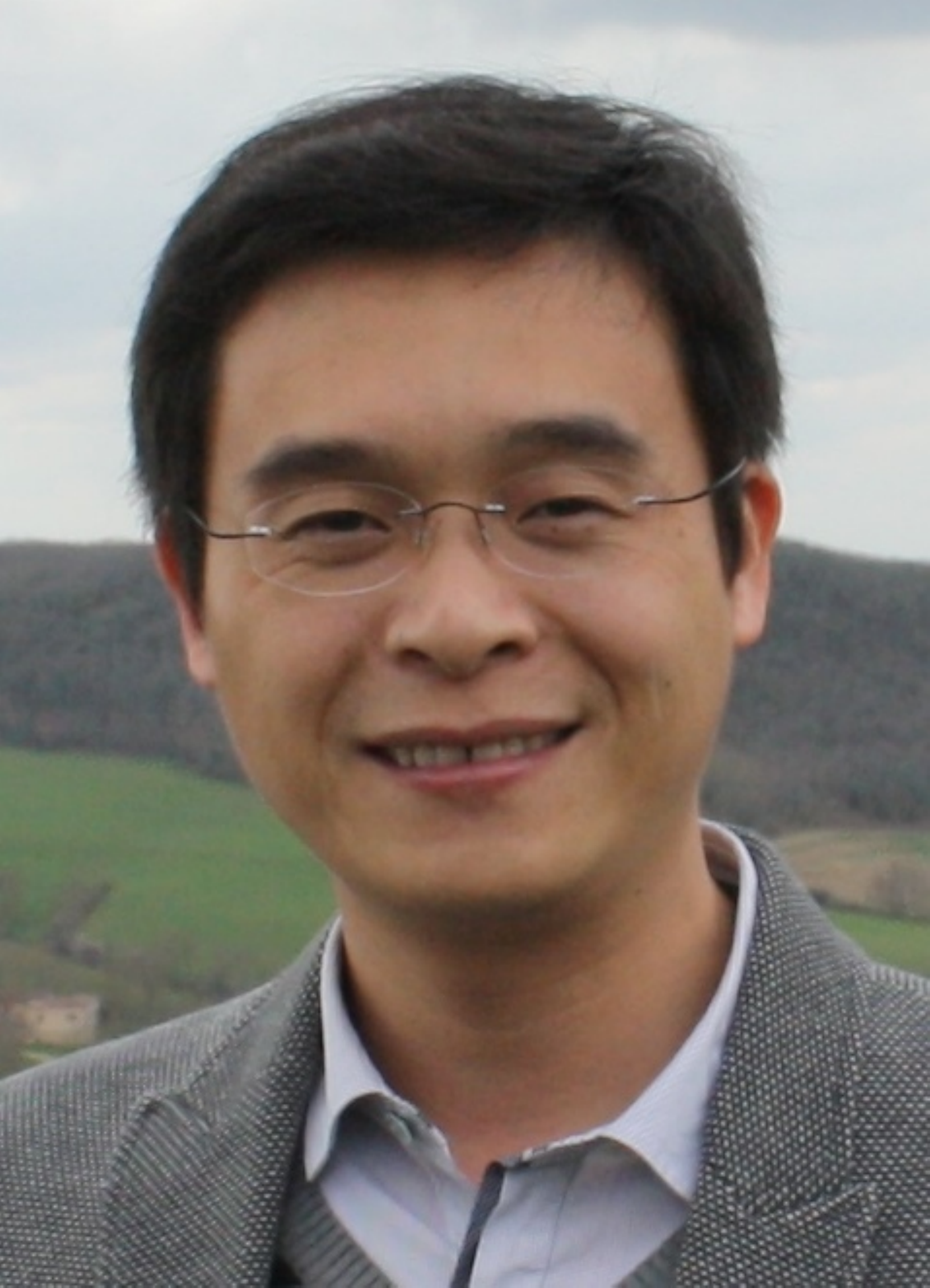}}]{Yongwang Zhao}
received the Ph.D. degree in computer science from Beihang University (BUAA) in Beijing, China, in 2009. He is an associate professor  at the School of Computer Science and Engineering, Beihang Univerisity. He has also been a Research Fellow in the School of Computer Science and Engineering, Nanyang Technological University, Singapore, from 2015. His research interests include formal methods, OS kernels, information-flow security, and AADL. 
\end{IEEEbiography}

\vspace*{-4\baselineskip}
\begin{IEEEbiography}[{\includegraphics[width=1in,height=1.25in,clip,keepaspectratio]{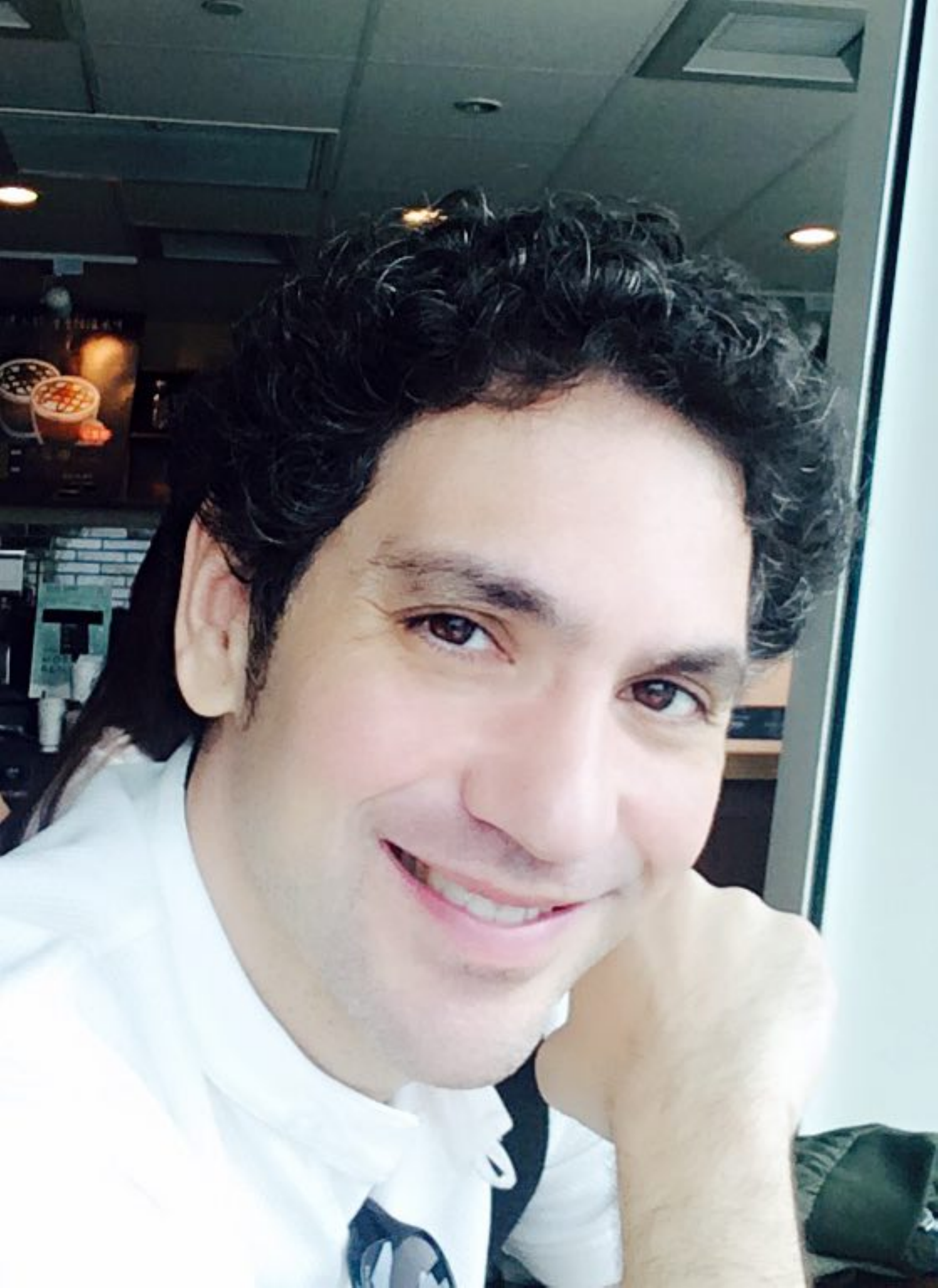}}]{David San\'an}
received the Ph.D. degree in Software Engineering and Artificial Inteligent from the University of M\'alaga, M\'alaga, Spain, in 2009. He has been working as a Research Fellow in the Singapore University of Technology and Design (SUTD), Trinity College Dublin (TCD), and National University of Singapore (NUS). In 2015 he joined Nanyang Technological University in Singapore, where he is currently working as a research fellow. His interest research includes formal methods, and in particular the formalization and verification of separation micro-kernels aiming multi-core architectures.
\end{IEEEbiography}

\vspace*{-4\baselineskip}
\begin{IEEEbiography}[{\includegraphics[width=1in,height=1.25in,clip,keepaspectratio]{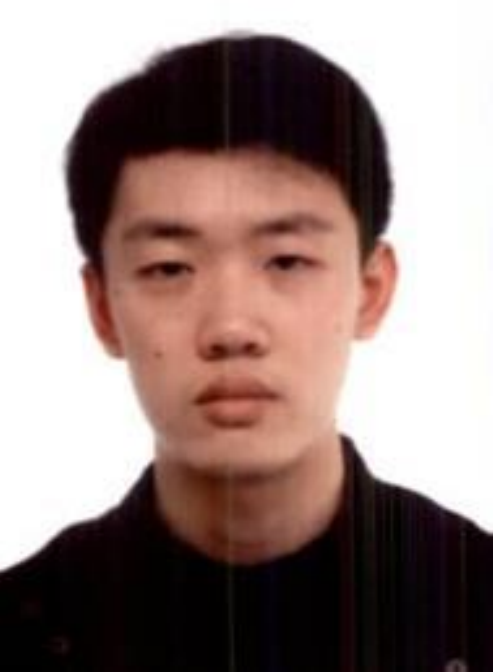}}]{Fuyuan Zhang}
received the Ph.D. degree in computer science from Technical University of Denmark in 2012. Currently, he is a research fellow in School of Physical and Mathematical Sciences, Nanyang Technological University, Singapore. His research interests include formal verification of IT systems, computer security and quantum computation.
\end{IEEEbiography}

\vspace*{-4\baselineskip}
\begin{IEEEbiography}[{\includegraphics[width=1in,height=1.25in,clip,keepaspectratio]{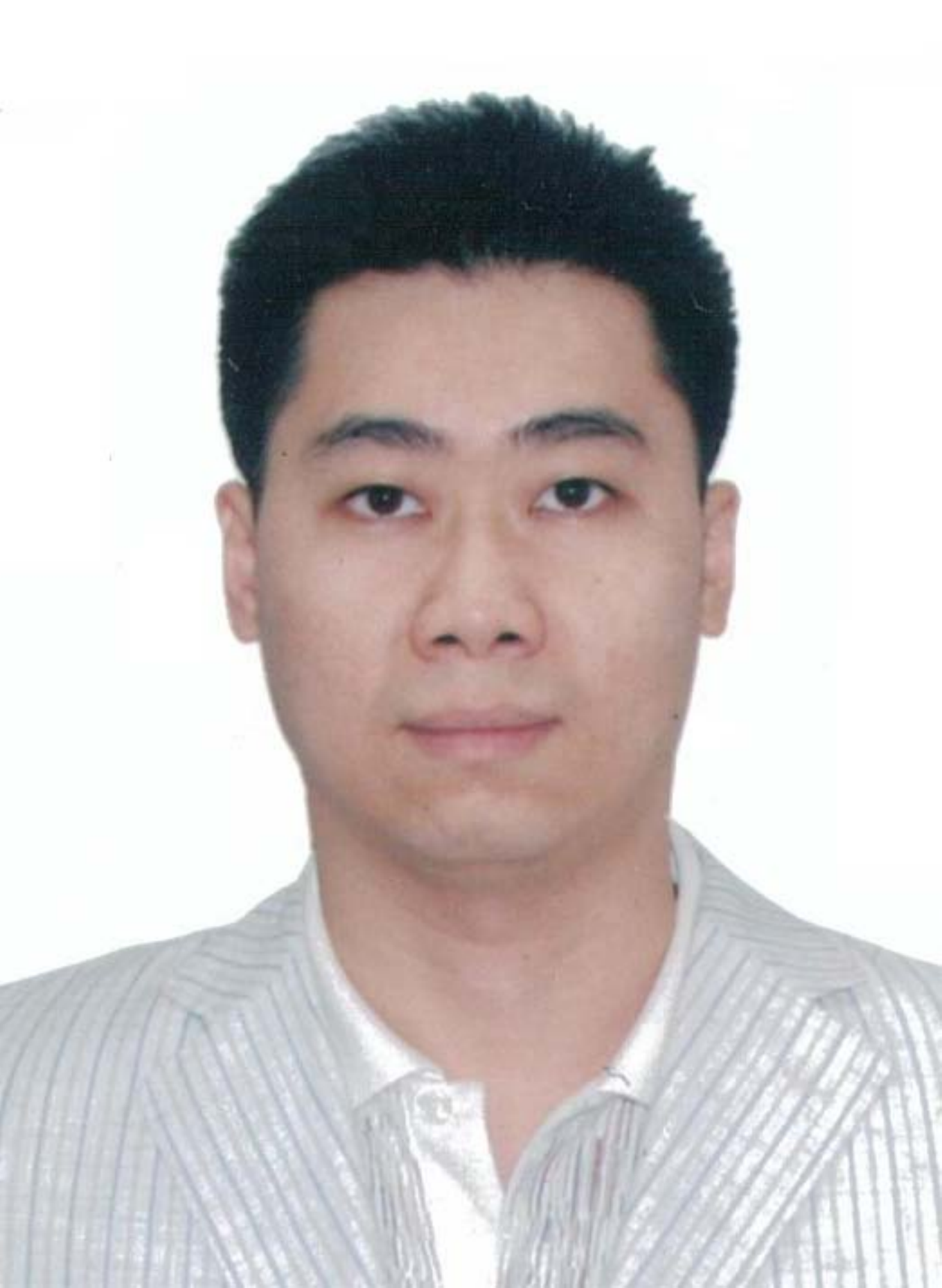}}]{Yang Liu}
received the Ph.D. degree in computer science from National University of Singapore (NUS) in 2010 and continued with his post-doctoral work in NUS. Since 2012, he joined Nanyang Technological University as an Assistant Professor. His research focuses on software engineering, formal methods and security. Particularly, he specializes in software verification using model checking techniques. This work led to the development of a state-of-the art model checker, Process Analysis Toolkit.
\end{IEEEbiography}

\appendices

\section{Covert Channels in ARINC 653}
\label{app:covt_ch_arinc}

The covert channels in ARINC 653 are discussed and fixed in our specification as follows. 

\textbf{Covert Channel 1 - Queuing mode channel}: 
If there is a queuing mode channel from a partition $a$ to a partition $b$ and no other channels exist, then it is secure that $a \interf \transmitter$, $\transmitter \interf b$, $\transmitter \ninterf a$ and $b \ninterf \transmitter$ according to the security policies in this paper. 
Actually, these security policies are violated in ARINC 653.
First, when $a$ sends a message by invoking the \emph{Send\_Queuing\_Message} service of ARINC 653, the service returns \emph{NOT\_AVAILABLE} or \emph{TIMED\_OUT} when the buffer is full, and returns \emph{NO\_ERROR} when the buffer is not full. However, the full/empty status of the buffer in the port can be changed by message transmission executed by the transmitter. Thus, the \emph{locally respects} condition is not preserved on the event of message transmission, and $\transmitter \ninterf a$ is violated. Second, due to no message loss allowed in ARINC 653, the transmitter cannot transmit a message on a channel when the destination queuing port is full. However, the full status of the destination port can be changed by the \emph{Receive\_Queuing\_Message} service executed by the partition $b$. Thus, the \emph{locally respects} condition is not preserved on \emph{Receive\_Queuing\_Message}, and $b \ninterf \transmitter$ is violated. To avoid this covert channel, we allow message loss on queuing mode channels in our specification.

\textbf{Covert Channel 2 - Leakage of port identifiers}:
It is assumed in ARINC 653 that a port identifier is only stored in a partition after creation. In the \emph{Send/Receive\_Queuing\_Message} and \emph{Write/Read\_Sampling\_Message} services, there is no judgement on whether the accessing port belongs to the current partition. 
Thus, the \emph{locally respects} condition is not preserved on the events. 
In this case, programs in a partition can guess the port identifiers of other partitions and then manipulate the ports. Therefore, the port identifier in ARINC 653 is a covert channel. This covert channel is avoided in our specification by checking that the port belongs to the current partition. 

\textbf{Covert Channel 3 - Shared space of port identifiers}:
In ARINC 653, the \emph{Create\_Sampling\_Port} and \emph{Create\_Queuing\_Port} services create a port and return a new unique identifier assigned by the kernel to the new port. Careless design of the port identifier can cause covert channels. In the initial specification, we use a natural number to maintain this new identifier. This number is initially assigned to ``1'' and increased by one after each port creation. In such as design, the two events do not preserve the \emph{step consistent} condition. Thus, the number becomes a covert channel that can flow information from any partition to others. This covert channel is then avoided in our specification by assigning the port identifier to each port during system initialization or in the system configuration. 

\textbf{Covert Channel 4 - Partition scheduling}:
ARINC 653 defines a cyclic partition scheduling. The time windows of a partition in the \emph{IDLE} mode are not preempted by other partitions. The result of executing the \emph{Schedule} event is that the current executing partition is set to a partition or the message transmitter. Although some partition is in \emph{IDLE} mode, it is also possible to be selected. That means the execution of \emph{Schedule} is independent with the state of partitions. 
However, state related scheduling policies, such as policies only selecting non-\emph{IDLE} partitions, do not have the information-flow security. In such a case, the \emph{step consistent} condition is not preserved on the \emph{Schedule} event, and the \emph{Set\_Partition\_Mode} service in partitions can interfere with the scheduler. This covert channel can be avoided by disabling the insecure partition scheduling. 

\textbf{Covert Channel 5 - Shared space of process identifiers}:
In ARINC 653, the \emph{Create\_Process} service creates a process and assigns an unique identifier to the created process. 
Careless design of the process identifier can cause covert channels. In the initial specification, we use global identifiers. When creating a process, we assign a new identifier which is not used by created processes. In such as design, the \emph{Create\_Process} service does not preserve the \emph{step consistent} condition. Thus, the global identifiers become covert channels that can flow information from any partition to others. This covert channel is then avoided in our specification by assigning a new identifier which is not used by created processes in current executing partition. Unlike communication ports, processes in ARINC 653 are not configured at built-time. 

\textbf{Covert Channel 6 - Leakage of process identifiers}:
It is assumed in ARINC 653 that a process identifier is only stored in a partition after creation. In process management services, e.g. \emph{Start/Stop/Suspend/Resume\_Process}, there is no judgement on whether the process belongs to the current partition. 
Thus, the \emph{locally respects} condition is not preserved on the events. 
In such a case, programs in a partition can guess the process identifiers of other partitions and then manipulate them. Therefore, created processes in ARINC 653 become covert channels. This covert channel is avoided in our specification by checking that the process belongs to the current partition.

\section{Covert Channels in Implementations} 
\label{app:covt_ch_impls}

We find five covert channels in VxWorks shown as follows. The covert channel 5 is not found during code review. 

\textbf{Covert channel 1}: In VxWorks 653, the message transmission is not implemented as a system event, but invoked by the \emph{Send\_Queuing\_Message} service. If there is a queuing mode channel $c$ from a partition $a$ to a partition $b$ and no other channels exist, then it is secure that $a \interf b$ and $b \ninterf a$. VxWorks uses the \emph{portQMsgPut} and \emph{portQMsgGet} functions to implement the \emph{Send\_Queuing\_Message} and \emph{Receive\_Queuing\_Message} services, respectively. According to the source code of \emph{portQMsgPut}, we find that when partition $a$ sends messages and the source port of $c$ is full, it invokes the \emph{portQMsgDistribute} function to transmit messages in the source port to the destination port. But when the destination port is also full, the \emph{portQMsgPut} function returns an error. 
However, the full status of the destination port can be changed by the \emph{portQMsgGet} function executed in partition $b$. 
Thus, the \emph{locally respects} condition is not preserved on \emph{portQMsgGet}, and covert channel 1 exists in VxWorks 653. 

\textbf{Covert channel 2}: VxWorks 653 uses the \emph{portSMsgPut} and \emph{portSMsgGet} functions to implement the \emph{Write\_Sampling\_Message} and \emph{Read\_Sampling\_Message} services, respectively. According to the source code of the \emph{portQMsgPut}, \emph{portQMsgGet}, \emph{portSMsgPut}, and \emph{portSMsgGet} functions, we find that when accessing a port, VxWorks 653 does not check that the port belongs to the current partition. Thus, these functions executed in a partition can manipulate ports in another noninterfering partition. The \emph{locally respects} condition is not preserved on these functions and covert channel 2 exists in VxWorks 653. 

\textbf{Covert channel 3}: VxWorks 653 uses the \emph{portQCreate} and \emph{portSCreate} functions to implement the \emph{Create\_Queuing\_Port} and \emph{Create\_Sampling\_Port} services, respectively. According to the source code of \emph{portQCreate} and \emph{portSCreate}, VxWorks 653 uses an object classifier to allocate port identifiers by invoking the \emph{objAlloc} function. When invoking \emph{objAlloc}, it uses the same object class (\emph{portQClassId}) for all queuing ports. Thus, the \emph{step consistent} condition is not preserved on the two functions and the \emph{portQClassId} becomes a covert channel, i.e. covert channel 3.

\textbf{Covert channel 4}: Beside the ARINC 653 partition scheduling, in which the partition scheduler is not interfered by partitions, VxWorks also supports Priority Pre-emptive Scheduling (PPS) for partitions. The PPS runs ARINC partitions in a priority pre-emptive manner during idle time in an ARINC schedule. The PPS is implemented in the \emph{ppsSchedulePartition} function, which choose the partition with the highest priority and ready tasks. In such a scheduling, the \emph{step consistent} condition is not preserved on the \emph{ppsSchedulePartition} function, and the \emph{Set\_Partition\_Mode} service and the services of process management in partitions can interfere with the scheduler. Thus, covert channel 4 potentially exists.

\textbf{Covert channel 6}: VxWorks 653 uses the \emph{taskActivate}, \emph{taskStop}, \emph{taskSuspend}, and \emph{taskResume} functions to implement the \emph{Start\_Process}, \emph{Stop\_Process}, \emph{Suspend\_Process}, and \emph{Resume\_Process} services, respectively. In these functions of VxWorks 653, it does not check that the accessing task belongs to the current partition. Thus, the \emph{locally respects} condition is not preserved on the functions. 
In such a case, programs in a partition may manipulate tasks in other partitions that the partition can not interfere with. Therefore, created tasks in VxWorks 653 become covert channels, i.e. covert channel 6.

We find two covert channels in XtratuM as follows. During code review of unwinding conditions, other covert channels are not found.

\textbf{Covert channel 1}: XtratuM uses one shared buffer between the source port and the destination port of a queuing mode channel as a transmitter. If there is a queuing mode channel $c$ from a partition $a$ to a partition $b$ and no other channels exist, then it is secure that $a \interf b$ and $b \ninterf a$. XtratuM uses the \emph{SendQueuingPort} and \emph{ReceiveQueuingPort} hypercalls to implement the \emph{Send\_Queuing\_Message} and \emph{Receive\_Queuing\_Message} services, respectively. According to the source code of \emph{SendQueuingPort}, we find if the buffer is not full, the hypercall \emph{SendQueuingPort} inserts the message into the buffer and notifies the receiver; whilst if the buffer is full, \emph{SendQueuingPort} immediately returns \emph{XM\_OP\_NOT\_ALLOWED}. However, the full status of the buffer can be changed by the \emph{ReceiveQueuingPort} hypercall executed in partition $b$. Thus, the \emph{locally respects} condition is not preserved on \emph{ReceiveQueuingPort}, and covert channel 1 exists in XtratuM. 

\textbf{Covert channel 4}: Beside the ARINC 653 partition scheduling, in which the partition scheduler is not interfered by partitions, XtratuM also supports fixed priority partition scheduling (FPS). The FPS chooses the \emph{READY} partition with the highest priority to be executed. The \emph{Schedule} function in XtratuM could be configured at built-time as FPS or ARINC 653 scheduling. In the FPS, the \emph{step consistent} condition is not preserved on the \emph{Schedule} function, and the \emph{Set\_Partition\_Mode} service in partitions can interfere with the scheduler. Thus, covert channel 4 exists in XtratuM.

We find one covert channel in POK as follows. During code review of unwinding conditions, other covert channels are not found.

\textbf{Covert channel 1}:
POK has a transmitter to transfer messages from a source port to a destination port of a channel. POK blocks processes to wait for resources. 
If there is a queuing mode channel from a partition $a$ to a partition $b$ and no other channels exist, then it is secure that $a \interf \transmitter$, $\transmitter \interf b$, $\transmitter \ninterf a$ and $b \ninterf \transmitter$. 
POK uses the \emph{pok\_port\_queueing\_send} and \emph{pok\_port\_queueing\_receive} syscalls to implement the \emph{Send\_Queuing\_Message} and \emph{Receive\_Queuing\_Message} services, respectively. First, according to the source code of \emph{pok\_port\_queueing\_send}, we find if the buffer of the source port is not full, \emph{pok\_port\_queueing\_send} inserts the message into the buffer; whilst if the buffer is full and $timeout = 0$, it immediately returns \emph{POK\_ERRNO\_FULL}. However, the full status of the source port can be changed by the \emph{pok\_port\_transfer} function which is in charge of transmitting messages from a source port to a destination one. Thus, the \emph{locally respects} condition is not preserved on \emph{pok\_port\_transfer}, and $\transmitter \ninterf a$ is violated. Second, \emph{pok\_port\_transfer} returns \emph{POK\_ERRNO\_SIZE} when the destination port has no available space to store messages. However, the full status of the destination port can be changed by the \emph{pok\_port\_queueing\_receive} function executed by partition $b$. Thus, the \emph{locally respects} condition is not preserved on \emph{pok\_port\_queueing\_receive}, and $b \ninterf \transmitter$ is violated.

\section{Attack Analysis of Covert Channels}
\label{app:att_anal}

We illustrate potential attacks to {\skname}s as follows. 

\textbf{Covert channel 1}: It is a typical storage channel between a sender and a receiver \cite{NCSC93}. If there is a queuing mode channel from a partition $a$ to a partition $b$, malicious programs in $a$ and $b$ can collaborate to create a covert channel from $b$ to $a$. Partition $a$ sends messages to the channel until the buffer is full. Then, the \emph{receive} event in $b$ can be inferred by $a$, and thus $a$ gets one bit of information from $b$ at each time of \emph{receive}. By the covert channel, any secret information in $b$ can be sent to $a$. 

\textbf{Covert channel 2}: It is a storage channel about resource isolation and leakage in {\skname}s. Let's assume a system with three partitions ($a$, $b$, and $c$) and a communication channel $ch$ from $b$ to $c$. According to security policies, information in $a$ cannot be leaked to other partitions. However, by the covert channel, malicious programs in partition $a$ can manipulate the channel $ch$ and send secret information of $a$ to $c$. 

\textbf{Covert channels 3 and 5}: They are storage channels on the shared identifiers among partitions. Malicious programs in partitions have to create resources (ports and processes) to reach the maximum number of the identifiers, and then can send one bit of information. Thus, the covert channels have very low bandwidth in real-world systems. 

\textbf{Covert channel 4}: It is a timing channel by the PPS in VxWorks 653 and the FPS in XtratuM. Higher prioritised partitions can directly influence when and for how long lower prioritised partitions run \cite{Volp08}.

\textbf{Covert channel 6}: It is a storage channel about resource isolation and leakage in {\skname}s. Assume a system with a set of partitions and there is no communication channel between partitions $a$ and $b$. According to security policies, information in $a$ and $b$ has to be isolated. However, by the covert channel, malicious programs in $a$ can manipulate the processes (e.g. change the process state) in $b$, and thus send some bits of information to $b$ at each time.

\end{document}